\documentclass{article}
\usepackage{amsthm,amsmath,amssymb,graphicx,amsfonts,array,url,enumerate,wrapfig,subfigure,cite,enumerate,fullpage}
\newtheorem{theorem}{Theorem}[section]
\newtheorem{lemma}{Lemma}[section]

\newtheorem{remark}{Remark}[section]

\def \Sone{{{\cal S}^1}}
\def \Stwo{{{\cal S}^2}}
\def \bone{{\beta_1}}
\def \btwo{{\beta_2}}
\def \bS{{\beta_S}}

\def \SNR{{\text{SNR}}}

\def \So{{S_\omega}}

\def \dx{\, \mathrm{d}}

\begin{document}

\title{Sparse Recovery with Linear and Nonlinear Observations: Dependent and Noisy Data}

\author{Cem Aksoylar and Venkatesh Saligrama}

\date{\vspace{-2ex}}

\maketitle

\begin{abstract}
  We formulate sparse support recovery as a salient set identification problem and use information-theoretic analyses to characterize the recovery performance and sample complexity. We consider a very general model where we are not restricted to linear models or specific distributions. We state non-asymptotic bounds on recovery probability and a tight mutual information formula for sample complexity. We evaluate our bounds for applications such as sparse linear regression and explicitly characterize effects of correlation or noisy features on recovery performance. We show improvements upon previous work and identify gaps between the performance of recovery algorithms and fundamental information. 
\end{abstract}

\section{Introduction}

In this paper, we consider problems where among a set of $D$ variables/features $X = (X_1,\ldots,X_D)$, only $K$ variables (indexed by set $S$) are directly relevant to the observation/label $Y$. These types of problems frequently arise in a number of scenarios in high-dimensional analysis, such as compressive sensing \cite{donoho}, feature selection in learning \cite{pattern} or other high-dimensional problems with an inherent low-dimensional structure. We formulate these problems with the following Markovian property: Given $X_S = \{X_k\}_{k \in S}$, observation $Y$ is independent of other variables $\{X_k\}_{k \not\in S}$, i.e.,
\begin{equation} \label{eq:markov}
P(Y|X) = P(Y|X_S).
\end{equation}

Given $N$ sample pairs $(X^N, Y^N) = \{(X^{(1)}, Y^{(1)}), \ldots, (X^{(N)}, Y^{(N)})\}$, our goal is to identify the set of relevant/salient variables $S$ from these $N$ samples.  
Our analysis aims to characterize the recovery performance probabilistically and establish necessary \& sufficient conditions on $N$ in order to recover the set $S$ with an arbitrarily small error probability in terms of $K$, $D$ and model parameters such as the signal-to-noise ratio (SNR). 

As an illustrative example, consider the sparse linear regression model, given by $Y^N = X^N \beta + W^N$, where $S$ is the support of sparse random vector $\beta$ and random noise $W^N$ assumed to be independent of $X^N$ and $\beta$. 
The elements in a row of the signal matrix correspond to variables $X_1,\ldots,X_D$. Each row is a realization of $X$ and $X^N$ is formed from sampled rows. Markov assumption \eqref{eq:markov} is satisfied, since each $Y^{(n)}$ depends only on the linear combination of the elements $X_S^{(n)}$ that correspond to the support of $\beta$. The coefficients of this combination are given by $\beta_S$, viewed as a random ``nuisance'' parameter in the observation model. This perspective also holds for non-linear models, thus unifying many sparse recovery problems. 

Information-theoretic approaches with relation to channel coding \cite{coverbook} have been considered in previous work for different application areas, where the salient set $S$ is seen as a message encoded by $X^N$ and is recovered from outputs $Y^N$. Specifically, the problem of group testing was formulated in a similar framework in Russian literature \cite{malyutov,malyutov1,malyutov2,malyutov3,dyachkov_lectures} and in \cite{group_testing}. \cite{itw} has followed a similar approach to \cite{group_testing} to obtain sample complexity results for general sparse signal processing models. Note that the identification problem formulated here has key differences with channel coding, namely the inability to ``code'' the variables $X^N$ and different messages/sets overlapping and thus sharing codewords.

\begin{figure}[h]
\centering
\includegraphics[width=0.45\textwidth]{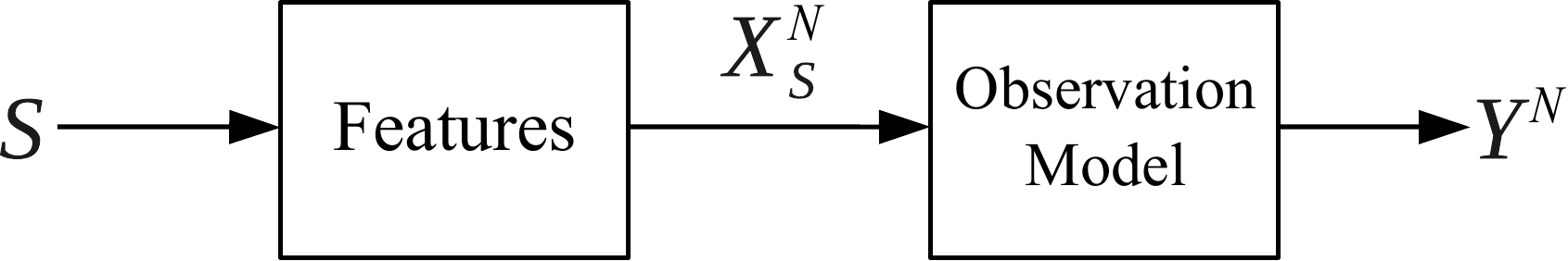} \caption{Channel model.}
\label{fig:channel_model}
\end{figure}

Previous work on general models, namely \cite{itw} has severe limitations related to the specific analysis techniques used. In this work we overcome several of these limitations, where we (1) consider correlated variables instead of independent and identically distributed (IID) variables and relate correlation to sample complexity, (2) present a non-asymptotic analysis through probability of recovery bounds instead of solely asymptotic analyses for sample complexity and (3) state more general results for the high-dimensional sparsity regime where sparsity scales with the number of variables. 

The bounds we present for the sample complexity are of the form
 \begin{equation} \label{eq:highlevel}
N \, I(X_S ; Y) > \log{\binom{D}{K}},
\end{equation}
which can be interpreted as follows: The right side of the inequality is the number of bits required to represent all sets $S$ of size $K$. On the left side, the mutual information term \cite{coverbook} represents the uncertainty reduction on the output $Y$ when given the input $X_S$, in bits per sample. This term essentially quantifies the ``capacity'' of the observation model $P(Y|X_S)$. \eqref{eq:highlevel} is then a statement that uncertainty reduction with $N$ samples should exceed uncertainty of set $S$.

Furthermore, our analysis also provides us with a sharp exponential upper bound on the probability of error in identifying the salient set. This bound can be computed easily and a closed form expression can be obtained for some applications, such as in the case of linear models. We compute these bounds for the sparse linear model described, where we explicitly characterize the effects of correlation, SNR or noisy variables.

Many models sharing the common structure of sparsity satisfy Markovian assumption \eqref{eq:markov}. These include sparse linear regression or compressive sensing (CS)~\cite{donoho}, probit regression or 1-bit CS~\cite{1bit,planvershinin}, group testing \cite{group_testing}, sparse logistic regression and multiple regression models~\cite{negahban} with group sparsity property. 
In addition, variants of these problems can be considered, e.g., with noisy or missing data where variables are not fully observed (see \cite{loh}).

\subsection{Related Work and Contributions}

The problem of sparse recovery and in particular information-theoretic (IT) analysis is extensive. We only describe work closely related to this paper. 
Much of the IT literature deals with linear models and mean-squared estimation of $\beta$ in the sparse linear model with sub-Gaussian assumptions on variables $X^N$. Below we list the contributions of our approach and contrast it with some of the related work in the literature. The Markovian problem formulation follows along the lines of \cite{itw,arxiv_iid}, which we repeat here for the sake of exposition.

{\bf Unifying framework through Markovianity:} 
Much of the literature on sparse recovery is specialized with tailored algorithms for different problems. For instance, lasso for linear regression \cite{candesplan,wainwright2},  relaxed integer programs for group testing \cite{jaggi}, convex programs for 1-bit quantization~\cite{planvershinin}, projected gradient descent for sparse regression with noisy and missing data~\cite{loh} and other general forms of penalization. While all of these problems share an underlying sparse structure, it is conceptually unclear from a purely IT perspective, how they come together from an inference point-of-view. 
Our Markovian viewpoint of \eqref{eq:markov} unifies these different sparse problems from an inference perspective. 

{\bf Discrete objects with continuous observations:} 
While \cite{wainwright, goyal, wainwright2, shuchin, akcakaya, yihong, reeves} describe IT bounds for sparsity pattern recovery to recover $S$, they exclusively focus on the linear sub-Gaussian setting. Furthermore, their approach is \emph{circuitous}. Indeed, they rely on first estimating the sparse vector $\beta$, which is then thresholded to obtain $S$. This not only complicates the analysis and introduces unnecessary assumptions on $\beta$ but also obfuscates the distinction between signal estimation vs.\ support discovery. It is well-known that if support is known, signal estimation is easy and least-squares estimates are reliable. At a conceptual level IT tools such as Fano's inequality and capacity theorems are powerful tools for inferring about discrete objects (messages) given continuous observations. Indeed, to exploit IT tools, \cite{wainwright, goyal, shuchin, akcakaya, yihong, reeves} resort to one of the following strategies: (a) Use IT tools only for necessity part by assuming a special case of discrete $\beta$ and derive sufficiency with some well-known algorithm (lasso, basis pursuit etc.); or (b) find a $\epsilon$-cover for $\beta$ in some metric space (which requires imposing extra assumptions) and reduce $\beta$ to a discrete object. In contrast our approach lifts these assumptions and focuses on the natural discrete object $S$. Our result shows that indeed the discrete part, namely, uncertainty support pattern is the dominating factor and not $\beta$ itself. 

Furthermore, prior work relied heavily on the design of sampling matrices with special structures such as Gaussian ensembles and RIP matrices, which is a key difference from the setting we consider herein as for our purpose we do not always have the freedom to design the matrix $X$. We do not make explicit assumptions about the structure of the sensing matrix, such as the restricted isometry property \cite{candes_rip} or incoherence properties \cite{candesplan}, or about the distribution of the matrix elements, such as sub-Gaussianity. Also, the existing IT bounds, which are largely based on Gaussian ensembles, are limited to the linear CS model, and hence not suitable for the non-linear models we consider herein.

{\bf Information-theoretic tight error bounds:} 
Through our analysis of the ML decoder, we obtain a tight upper bound on the probability of error of support recovery, in addition to necessary and sufficient conditions on the sample complexity. We compute this upper bound explicitly for popular problems such as sparse linear regression and its variants. 
We compare the information-theoretic bound to the performance of practical algorithms used to solve the sparse recovery problem, such as lasso \cite{candesplan, wainwright2} or orthogonal matching pursuit (OMP) variants \cite{caramanis} and illustrate large gaps between their performance and our bounds. The presence of these gaps show that there is still room to improve the performance of practical algorithms for solving support recovery problems. 

{\bf Bounds for new sparse recovery problems:} 
Our unifying approach also allows the study of problems that are not previously analyzed, or that are not easily analyzed using other approaches. These types of problems may include sparse recovery with novel observation models, or existing models with different distributions of variables or noise. Due to our Markovian formulation, obtaining necessary and sufficient conditions and error bounds only necessitate computation of simple mutual information expressions and an error exponent expression.

As mentioned in the introduction, the identification problem was formulated in a channel coding framework in \cite{group_testing} and in the Russian literature \cite{malyutov,malyutov1,malyutov2,malyutov3,dyachkov_lectures}. 
This analysis was extended to general sparse signal processing models with IID variables and latent variable observation model in \cite{itw} and \cite{arxiv_iid}. 
 
We consider the analysis of models with correlated variables, specifically conditionally IID variables $X$ given a latent parameter $\theta$. We also state a non-asymptotic bound on the probability of error, which in turn allows us to identify performance gaps between practical algorithms and our information-theoretic results. In addition, we consider a general scaling regime where $K = O(D)$ for linear models and variants through this bound. 
We also introduce the noisy data framework and explicitly characterize the recovery performance w.r.t.\ the noise variance.

\subsection{Problem setup}
\label{sec:problem}

\textbf{Notation.} 
We represent variables with row vectors and samples as different rows to obtain a $N \times D$ variable matrix, while the observation samples form a column vector. In that context, subscripts are used for column indexing and superscripts with parentheses are used for row indexing in vectors and matrices. $\log$ is used to denote logarithm to the base $2$.

\textbf{Problem setup.} We observe the realizations of $N$ variable-observation pairs $(X^N,Y^N)$ with each sample $(X^{(n)},Y^{(n)})$, $n=1,2,\ldots,N$. Observations $Y$ are given by $P(Y|X_S,\beta_S)$ with latent model parameter $\beta_S \sim P(\beta_S)$ and satisfy the Markovian property \eqref{eq:markov}, where $|S| \leq K$ with known $K \ll D$. Observation parameters $\beta_S$ correspond to the coefficients on the support of the sparse vector in sparse recovery problems. For simplicity of exposition we consider the case $|S| = K$. The variables $X^{(n)}$ are IID\ across $n=1,\ldots,N$. However, the observations $Y^{(n)}$ are independent for different $n$ only when conditioned on $\beta_S$. Our goal is to identify the set $S$ from the $N$ samples of variables and the associated observations $(X^N,Y^N)$, with an arbitrarily small average error probability. 

We index the different sets of size $K$ as $S_{\omega}$, so that $S_{\omega}$ is a set of $K$ indices corresponding to the $\omega$-th set of variables. Since there are $D$ variables in total, there are $\binom{D}{K}$ such sets, hence $\omega\in\left\{1,2,\ldots\binom{D}{K}\right\}$. 

Let $\hat{S}(X^N,Y^N)$ denote the estimate of the set $S$ which is random due to the randomness in $X$ and $Y$ and let $P(E)$ denote the average probability of error, averaged over all sets $S$ of size $K$, variables $X^N$ and observations $Y^N$, i.e., $P(E) = \Pr[\hat{S}(X^N,Y^N)\ne S] $.

\section{Recovery and Error Bounds}
\label{sec:recovery}

Central to our analysis are the following four assumptions, which we utilize in order to analyze the probability of error in recovering the salient set and to obtain bounds on sample complexity.

\begin{enumerate}[({A}1)]
\item 
\textbf{Equiprobable support:} Any set $S_\omega \subset \{1,\ldots,D\}$ with $K$ elements is equally likely \emph{a priori} to be the salient set.
We assume no prior knowledge of the salient set $S$ among $\binom{D}{K}$ such sets. 

\item
\textbf{Conditional independence:} The observation/label $Y$ is conditionally independent of other variables given $X_S$, variables with indices in $S$, i.e.,
$P(Y | X) = P(Y | X_S)$. 

\item
\textbf{Conditionally IID variables:} The variables $X_1, \ldots, X_D$ are IID conditioned on a latent parameter $\theta$. We elaborate on this property in the following sections. 

\item
\textbf{Observation model symmetry:} For any permutation mapping $\pi$, $P(Y | X_S) = P(Y | X_{\pi(S)})$, i.e., the observations are independent of the ordering of variables. 
This is not a very restrictive assumption since the asymmetry w.r.t.\ the indices can be incorporated into $\bS$, as the symmetry is assumed for the observation model averaged over $\bS$.

\end{enumerate}

With only these four general assumptions, we are able to identify bounds that we state in the next section, for a general class of problems. We now elaborate on some of the assumptions above. 

\subsubsection*{Conditional Independence}

A simple example with the conditional independence property is the sparse linear model mentioned in the introduction. In this model every observation is given by the model 
\[ Y = \langle X, \beta \rangle + W = \langle X_S, \beta_S \rangle + W \]
with noise $W$, which can also be extended to nonlinear models $Y = f(\langle X_S, \beta_S \rangle + W)$, for a function $f: \mathbb{R} \to \mathbb{R}$. 

A different nonlinear example is the group testing model, which is detailed in \cite{group_testing, arxiv_iid}. In this model observations are Boolean test results, while the variables are Boolean test inclusion indicators for each test and item. Each test result is positive if and only if any item from a certain defective set of items is included in that test, i.e.,
\[ Y = \bigvee_{k \in S} X_k. \]
Denoting the set of defective items with $S$, it is then clear that the conditional independence holds. 

\subsubsection*{Conditionally IID Variables}

For conditionally IID variables, the joint distribution of variables can be written as
\begin{equation*}\label{eq:cond_iid} 
P(X_1,\ldots,X_D) = \int_\Theta \prod_{k=1}^D P(X_k | \theta) P(\theta) \dx \theta, 
\end{equation*}
where $\theta \in \Theta$ is the latent coupling parameter with density $P(\theta)$. 
(A3) appears restrictive and so we describe a few examples and extensions to build intuition. 

{\bf Bouquet model \cite{bouquet}} arises in sparsity-based face recognition and given by 
$X_k = \mu + W_k, ~~~k=1,\ldots,D$, with $W_k \sim {\cal N}(0, \sigma_W^2)$ IID across $k$ and $\mu \sim {\cal N}(0, \sigma_\mu^2)$. It can be seen that two variables $X_i$ and $X_j$ are dependent and correlated with correlation coefficient $\rho = \sigma_\mu^2/(\sigma_\mu^2 + \sigma_W^2)$ but IID conditioned on $\mu$. 

{\bf Meta parameters:} We can account for several possibilities by selectively introducing meta parameters. For instance, we can let $X_k = \alpha_k^\top \mu$ with $\mu \sim {\cal N}(0,I_D)$ and IID random vectors $\alpha_k$. Here $(\{X_k\} \mid \{\alpha_k\}, \mu)$ are 
independent, not identically distributed with $E(X_k X_j \mid \{\alpha_k\}, \mu) = \alpha_k^\top \alpha_j$. Nevertheless, our results also extend to this setting. Note that $X_k$'s are exchangeable. Indeed, there is a close connection between conditional IID random variables and exchangeable random variables through de Finetti's theorem \cite{diaconis-freedman,aldous,lauritzen}. 

\subsection{Recovery Conditions and Error Bounds}
\label{subsec:recovery}

To derive the upper bound on recovery error and sufficiency bound for the required number of samples, we analyze the error probability of a Maximum Likelihood (ML) decoder \cite{gallager}. The decoder goes through all $\binom{D}{K}$ sets of size $K$ and chooses the set $S_{\omega^*}$ for which observation $Y^N$ is most likely, i.e.,
\begin{equation}
P(Y^N|X^N_{S_{\omega^*}}) > P(Y^N|X^N_{S_\omega}),~~~~\forall \omega\ne \omega^*.
\end{equation}
An error occurs if any set other than the true set $S$ is more likely. This ML decoder is a minimum probability of error decoder assuming uniform prior on the candidate sets of variables. Note that the ML decoder requires the knowledge of the observation model $P(Y|X_S,\beta_S)$ and the prior $P(\beta_S)$. Next, we derive an upper bound on the error probability $P(E)$ of the ML decoder, averaged over all sets, data realizations and observations.

Our methodology for the analysis is as follows: To deal with scenarios where a candidate set $\So$ and true set $S$ have overlapping elements (and thus $X_\So^N$ and $X_S^N$ share certain columns), we define the error event $E_i$ as the event of mistaking the true set for a set which differs from the true set $S$ in exactly $i$ variables, i.e., there exists some set which differs from the true set in $i$ variables and is more likely to the decoder. Note that $E = \bigcup_{i=1}^K E_i$. Then for each $i$ we use an analysis based on the characterization of error exponents as in \cite{gallager} to obtain an upper bound on $P(E_i)$, which leads to Theorem \ref{thm:P(E)} and a sufficient condition on $N$. We derive a matching necessity bound on $N$ with an argument based on Fano's inequality \cite{coverbook}.

Our first main result is the following theorem, which states a non-asymptotic upper bound on the probability of error of exact support recovery.

\begin{theorem} \label{thm:P(E)}
Under the assumptions (A1)-(A4), the probability of error $P(E)$ that a set other than $S$ is selected by the ML decoder 
is bounded from above by
\begin{equation}
P(E)\leq \min_{\delta \in [0, 1]} \sum_{i=1}^K 2^{-N\left(E_o(\delta)-\delta\frac{\log\binom{D-K}{i}\binom{K}{i}}{N}\right)},
\label{eq:error_exp_thm}
\end{equation}
where
\begin{align*}
  E_o (\delta) & = -\frac{1}{N} \log E_{\theta^N} \left[ \sum_{Y^N} \sum_{X_\Stwo^N} P(X_\Stwo^N | \theta^N) 
  \left(\sum_{X_\Sone^N} P(X_\Sone^N | \theta^N) P(Y^N|X_\Sone^N,X_\Stwo^N)^{\frac{1}{1+\delta}}\right)^{1+\delta} \, \right],
\end{align*}
for $0\leq\delta\leq 1$. 
$({\cal S}^1,{\cal S}^2)$ denotes a disjoint partition of the true set of variables $S$ with cardinalities $i$ and $K-i$, $X_\Sone^N$ and $X_\Stwo^N$ are the corresponding disjoint partitions of the $N\times K$ input $X_{\cal S}^N$ of sizes $N\times i$ and $N \times (K-i)$, respectively. $\theta$ is the parameter in the cond.\ IID representation. The bound holds for any $(N, K, D)$.
\end{theorem}

\begin{remark}
For fixed and known $\bS$, observations $Y^{(n)}$ are independent and $E_o(\delta)$ simplifies to
\begin{align*}
 E_o(\delta) = & - \log E_{\theta} \left[ \sum_Y \sum_{X_\Stwo} P(X_\Stwo | \theta) 
 \left( \sum_{X_\Sone}P(X_\Sone | \theta) P(Y | X_\Sone, X_\Stwo)^{\frac{1}{1+\delta}} \right)^{1+\delta} \right].
\end{align*}
\end{remark}

Next, we state our main result for the sample complexity of support recovery. The following theorem provides tight necessary and sufficient conditions on the number of samples $N$ asymptotically for an arbitrarily small average error probability. 

\begin{theorem}
\label{thm:main_theorem}
Let $I(X_\Sone;Y | X_\Stwo,\beta_S, \theta)$ be the mutual information between $X_\Sone$ and $Y$ conditioned on $X_\Stwo$, $\beta_S$ and $\theta$. 
Under the assumptions (A1)-(A4), a necessary condition on the number of samples $N$ to recover $S$ with an arbitrarily small average probability is given by
\begin{equation} \label{eq:main_theorem}
N > (1+\epsilon) \max_{i=1,\ldots,K} \frac{\log\binom{D-K+i}{i}}{I(X_\Sone;Y | X_\Stwo,\beta_S, \theta)}.
\end{equation}
Furthermore, if $I(X_\Sone;Y | X_\Stwo,\beta_S, \theta) = \omega(1/\log D)$ for all $i=1,\ldots,K$, \eqref{eq:main_theorem} is also a sufficient condition. The necessary condition holds for all scalings $K = O(D)$ and $\epsilon = 0$, while the sufficiency bound holds for any fixed $K$ as $D \to \infty$ and $\epsilon > 0$ an arbitrary constant.
\end{theorem}

Note that the condition that $I(X_\Sone;Y | X_\Stwo,\beta_S,\theta) = \omega(1/\log D)$ is not restrictive, since typically the mutual information per sample depends on the number of salient variables $K$ and not on the total number of variables $D$ and we consider the regime where $K$ is fixed w.r.t.\ $D$ for the sufficient condition.

\textbf{IID variables.} 
For IID variables, the mutual information expression in the denominator is $I(X_\Sone;Y | X_\Stwo,\beta_S)$ and further reduces to $I(X_\Sone; Y | X_\Stwo)$ for fixed observation parameters $\beta_S$.

\textbf{Interpretation.} 
Intuitively, the condition in \eqref{eq:main_theorem} can be explained as follows: For each $i$, the numerator is the number of bits required to represent all sets $S_\omega$ with $K-i$ indices known beforehand. The denominator represents the information given by the output variable $Y$ about the remaining $i$ indices $\Sone$, given the subset $\Stwo$ of $K-i$ true indices. Hence, the ratio represents the number of samples needed to control $i$ support errors in $\Sone$ and maximization accounts for all possible support errors.

\textbf{Partial recovery.} 
As we analyze the error probability separately for $i=1,\ldots,K$ support errors in order to obtain the necessity and sufficiency results, it is trivial to determine conditions for \emph{partial} instead of \emph{exact} support recovery. By changing the maximization from over $i=1,\ldots,K$ to $i= \lfloor \alpha K \rfloor,\ldots,K$ in \eqref{eq:main_theorem}, the conditions to recover at least $(1-\alpha)K$ of the $K$ support indices can be determined.

\textbf{Support pattern recovery dominates support coefficient estimation.} 
In the proof of Theorem \ref{thm:main_theorem}, we show that $\beta_S$ being unknown with prior $P(\beta_S)$ induces a penalty term in the denominator given by $I(\beta_S ; X_\Sone^N | X_\Stwo^N, Y^N, \theta^N)/N$, compared to the case where support coefficients $\beta_S$ are fixed and known. We show that this term is always dominated by $I(X_\Sone;Y | X_\Stwo,\beta_S, \theta)$ provided a mild condition on the mutual information is satisfied, therefore does not affect the sample complexity asymptotically. This shows that recovering support while knowing the support coefficients is as hard as recovering with unknown coefficients, underlying the importance of recovering the support in sparse recovery problems.

\section{Applications}
\label{sec:apps}

In this section, using the result of Theorem \ref{thm:P(E)}, we provide explicit non-asymptotic upper bounds for the error probability for sparse linear models that may include correlations or noisy variables. We also state asymptotic sample complexity results using the error bounds and Theorem \ref{thm:main_theorem}. We then compare the information-theoretic error bounds we obtained with the recovery performance of practical algorithms.

\subsection{Sparse linear regression}
\label{subsec:cs_bounds}

We consider the normalized model \cite{shuchin},
\begin{equation} \label{eq:cs_model}
  Y^N = X^N \beta + W^N, 
\end{equation}
where $X^N$ is the $N \times D$ sensing matrix, $\beta$ is a $K$-sparse vector of length $D$ with support $S$ and $Y^N$ is the observation vector of length $N$. We assume $X^{(n)}$ are jointly Gaussian row vectors and IID across rows $n$. Each element $X_k^{(n)}$ is zero mean and has variance $1/N$.
$W^N$ is the IID observation noise, with $W \sim {\cal N}\left(0, \frac{1}{\SNR}\right)$. The coefficients of the support, $\beta_S$, are either fixed and $|\beta_k| = \sigma$, or IID Gaussian with zero mean and variance $\sigma^2$.

\begin{wrapfigure}{R}{0.45\textwidth}
\vspace{-20pt}
\centering
\includegraphics[width=0.45\textwidth]{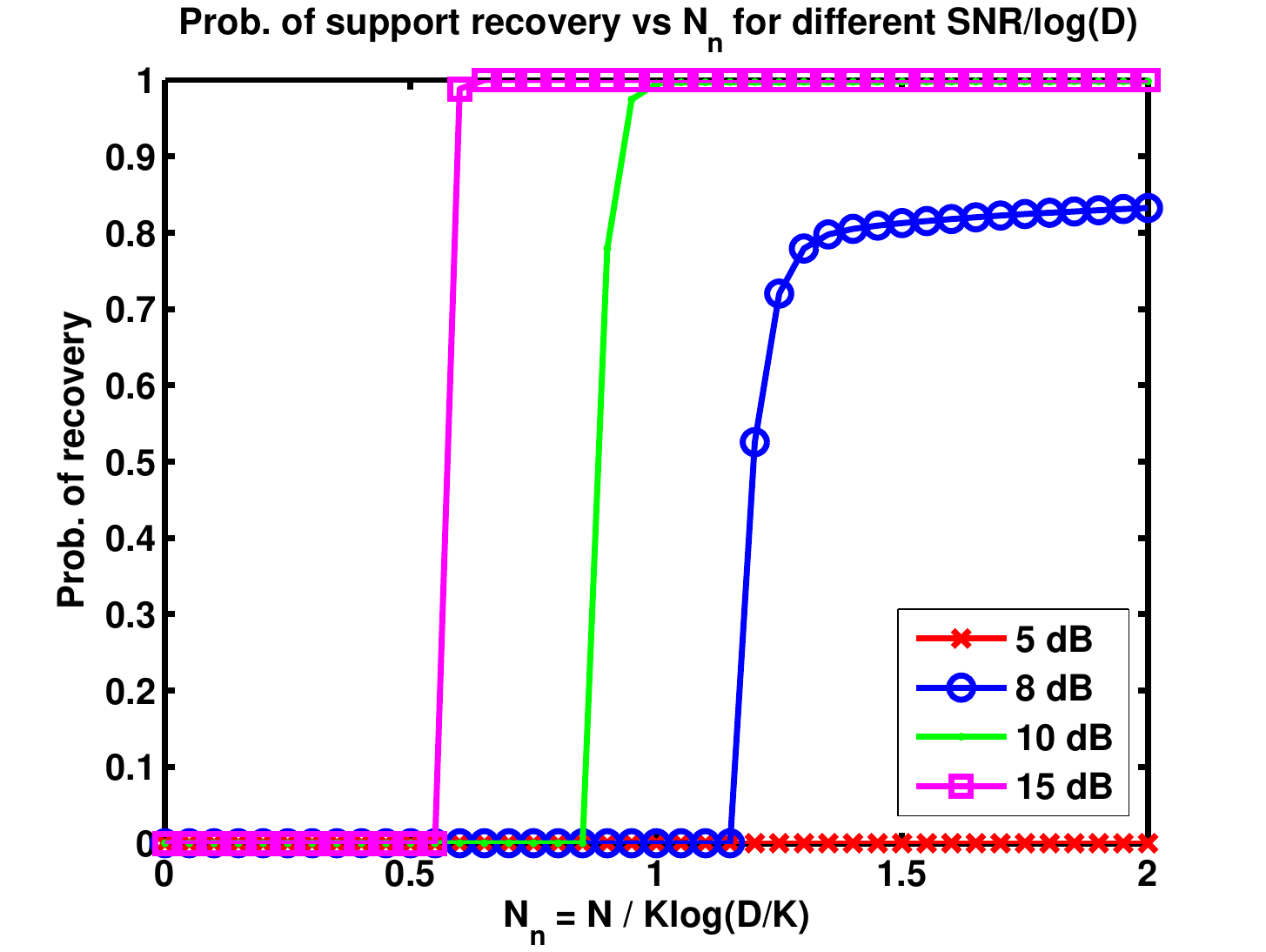}
\caption{Illustration of SNR cutoff, $K=32$, $D=512$.}
\label{fig:cs_SNR}
\vspace{-10pt}
\end{wrapfigure}

\begin{theorem}
\label{thm:lcs}
Consider the correlated setup described above. First, $\SNR = \Omega(\log D)$ is a necessary condition for recovery. Furthermore, for this SNR we can recover $S$ with average error probability approaching zero if and only if $N = \Omega\left(\frac{K \log (D/K)}{\log(1+(1-\rho)\sigma^2)}\right)$.
\end{theorem}
We consider a generalized model, which may include correlations between sensing columns, such that $E[X_k^{(n)} X_{k^\prime}^{(n)}] = \rho / N$. 
$\rho$ is then the correlation coefficient between two columns. Note that this model is statistically equivalent to the following one: Let $X_k^{(n)} = \mu^{(n)} + U_k^{(n)}$, where $\mu$ is also a Gaussian random variable with zero mean and variance $\rho/N$. 
$U_k^{(n)}$ is IID Gaussian, 
with zero mean and variance $(1-\rho)/N$. We analyze the latter model, where entries are conditionally IID given $\mu$. Correlated columns have been analyzed for lasso in this context~\cite{candesplan,wainwright2}. The strongest results due to \cite{candesplan} require correlations to decay asymptotically to zero as $1/\log(D)$, while \cite{wainwright2} is not strictly comparable since their results are for high-SNR limit. In contrast, we will show that fundamentally, up to constant correlation can be tolerated. The following theorem provides an upper bound to the probability of error for exact support recovery.

\begin{theorem} \label{thm:cs_bound}
$P(E) \leq \sum_{i=1}^K 2^{-N f(\rho) } $, 
where $f(\rho) = \frac{1}{2} \log \left( 1 + (1-\rho) \frac{2 i \sigma^2 \SNR}{N}\right)
  - \frac{i}{4N} \log 4 - \frac{\log\binom{D-K}{i}\binom{K}{i}}{N} 
$.
\end{theorem}

The first term in $f(\rho)$ is related to the information between $X$ and $Y$ via $\SNR$ and $\rho$, while the second term is related to the uncertainty of $\bS$ and the last term to the uncertainty of $S$. Note that the error bound given above precisely characterizes the achievable error for any $(N, K, D)$, in contrast to the setting for the sufficient condition in Section \ref{subsec:recovery} where $K$ is fixed w.r.t.\ $D$. Using this bound, we show the probability of recovery vs.\ other interesting quantities (see Figs. 4, 5). Also note the relation between $f(\rho)$ and $\rho$, e.g., for the degenerate case where $\rho = 1$, $f(\rho)$ is negative for any $N$. This is expected since recovery is not possible in that case, which we prove with the necessity bound.

We now present necessary and sufficient conditions for exact support recovery. We start with a lemma describing the mutual information for this model. 

\begin{lemma} \label{thm:cs_I}
\[ I(X_\Sone ;Y |  X_\Stwo, \beta_S, \mu) = \frac{1}{2} E \left[ \log \left( 1 + (1-\rho) \frac{\|\beta_\Sone\|^2  \SNR}{N} \right) \right],  
\]
where the expectation is w.r.t.\ ${\beta_\Sone}$.
\end{lemma}

The mutual information formula along with the bound given by Theorem \ref{thm:cs_bound} allow us to obtain the following necessary \& sufficient condition for exact recovery.

The necessary condition on SNR is also illustrated in Figure \ref{fig:cs_SNR}, where we plot the probability of error bound given by Theorem \ref{thm:cs_bound} for different SNR values. Indeed, we show an SNR cutoff regardless of number of measurements as well as tradeoffs beyond the cutoff point. Note that the relation between SNR and $N$ is not explicitly described for lasso \cite{wainwright2, candesplan}. 

Both upper and lower bounds hold for the general case $K = O(D)$, since we use the error bound in Theorem \ref{thm:cs_bound} to obtain the upper bound instead of Theorem \ref{thm:main_theorem}.

\begin{remark}
It follows that our relatively simple analysis gives us a bound asymptotically identical to the best-known bound $N = \Omega(K \log (D/K))$ \cite{shuchin} with an independent Gaussian sensing matrix. Our results also incorporate correlations to explicitly characterize the effect of correlated columns on sample complexity. We have shown that the number of samples increases by $\frac{1}{\log(1+(1-\rho)C)}$ relative to $\frac{1}{\log(1+C)}$ for the independent model for some constant $C$.
\end{remark}

\subsection{Noisy variables}
\label{subsec:cs_noisy_bounds}

We also analyze the additive noise model considered in \cite{loh,caramanis}, where in the sparse linear regression model \eqref{eq:cs_model}, a matrix $Z^N$ is observed instead of the sensing matrix $X^N$, with the relation
$Z^N = X^N + V^N$, 
where $V^{(n)} \sim {\cal N}(0, \frac{\nu}{N} I_D)$ IID for $n=1,\ldots,N$. The rest of the setup is as given in Section \ref{subsec:cs_bounds}. Note that in contrast to Section \ref{subsec:cs_bounds}, the model described here exhibits a nonlinear relationship between the variables $Z^N$ and the observations $Y^N$.

For this problem with noisy observations of variables, we have the following theorem for an upper bound on the probability of error of exact support recovery. 
\begin{theorem} \label{thm:cs_additive_bound}
The error probability of exact support recovery in the noisy data model is given by 
$P(E) \leq \sum_{i=1}^K 2^{-Nf(\rho,\nu)}$, 
where 
$f(\rho, \nu) = \frac{1}{2} \log \left( 1 + \frac{1-\rho}{1+\nu} \frac{2 i \sigma^2 \SNR}{N \xi} \right)
- \frac{i}{4N} \log 4 - \frac{\log\binom{D-K}{i}\binom{K}{i}}{N}$, where $\xi = 1 + \frac{(1-\rho)\nu}{1+\nu} \frac{K\SNR\sigma^2}{N}$. 
\end{theorem}

The error exponent $f(\rho, \nu)$ differs from $f(\rho)$ defined in Section \ref{subsec:cs_bounds} mainly by an extra $1 + \nu$ term in the denominator in the $\log$ term and reduces to $f(\rho)$ for $\nu = 0$. Also note that $\xi \approx 1$ for sufficiently large $N$.

We now state a sufficient condition on the number of measurements with the theorem below, which follows from an analysis of the upper bound on recovery error provided in Theorem \ref{thm:cs_additive_bound}.

\begin{theorem} \label{thm:cs_additive}
For $\SNR = \Omega\left(\log D\right)$, a sufficient condition on the number of measurements is $N = \Omega\left( \frac{K \log(D/K)}{\log\left( 1 + \frac{1-\rho}{1+\nu} \sigma^2 \right)} \right)$.
\end{theorem}

\begin{remark}
We observe that the sufficient number of measurements is affected by a factor of $\frac{1}{\log(1 + C/(1+\nu))}$ in our results, which greatly improves upon the bound with a factor of $(1 + \nu)^2$ by \cite{caramanis}.
\end{remark}

We also note that while \cite{wainwright, candesplan} analyze correlated Gaussians and others noisy or missing data \cite{loh, caramanis} separately, our error bounds and asymptotic sample complexity results unify these into a single expression.

\section{Experiments and Comparison with Achievable Bounds}
\label{sec:experiments}

In this section we compare the bounds on the probability of successful recovery we derived in the above sections with the frequency of successful exact support recovery for two recovery algorithms. 

For all experiments and evaluation of bounds, we set $K=32$ and $D=512$. The variables $X^N$ and observations $Y^N$ are generated according to the normalized model given by \eqref{eq:cs_model}, where we choose $S$ uniformly at random and let $\beta_S \in \{-1, 1\}^K$ with uniform probability. $N_n = N / (K \log(D/K))$ is the normalized number of measurements.

\subsection{Lasso and iteratively reweighted lasso}

We compare our bounds for independent and correlated sensing elements with lasso \cite{candesplan, wainwright2}, as defined in \cite{candesplan}. Formally, lasso gives the solution to the following optimization problem:
\[ \beta^\star = \arg \min_\beta \frac{1}{2} \left\| Y^N - X^N \beta \right\|_2^2 + \lambda \|\beta\|_1. \]

We set the regularization parameter as $\lambda = 2\sqrt{2 \log D}/\sqrt{\SNR}$ as suggested in \cite{candesplan}. We also investigated different values however we have not observed any significant improvements in performance. 

We also investigate the performance of a non-convex iterative lasso variant called the iteratively reweighted lasso. This method is proposed in \cite{reweighted} for the noiseless recovery problem; we use an extension for the noisy case, which iteratively solves the following optimization problem at each step:
\[ \beta^{(l)} = \arg \min_\beta \frac{1}{2} \left\| Y^N - X^N \beta \right\|_2^2 + \lambda_r \sum_{n=1}^N w_n^{(l)} |\beta_n|, \quad w_n^{(l)} = \frac{1}{\left|\beta_n^{(l-1)} \right| + \epsilon}. \]
This optimization is the same as lasso except for the individual weights $w_n^{(l)}$ for each component $\beta_n$, which depend on the output of the previous iteration. $\epsilon$ is a suitably small constant that stabilizes the weights for $|\beta_n|$ close to zero. Setting $\beta^{(0)}$ to the solution of regular lasso, the algorithm iterates until $\| \beta^{(l)} - \beta^{(l-1)} \|$ is smaller than a tolerance constant or a maximum number of iterations is reached.

Reweighted lasso aims to sparsify the estimated $\beta$ compared to regular lasso. At each iteration, it places greater weight on small variables to sparsify the solution, while the influence of large variables is reduced in order to allow for more sensitivity in identifying the other variables. The authors in \cite{reweighted} intuitively justify the sparsifying properties of the algorithm by noting that iteratively solving the reweighted $\ell_1$ problem is a Majorization-Minimization algorithm for the log-sum penalty problem, where the penalty is defined as $\sum_{n=1}^N \log\left(|\beta_n| + \epsilon\right)$. 
The sparsity encouraging properties of this method can be intuitively justified by the fact that $\log\left(|\beta_n| + \epsilon\right)$ approximates the $\ell_0$ penalty much better than $\ell_1$ does. It should be noted that the log-sum penalty is non-convex, therefore the iterative reweighted minimization is not guaranteed to converge to its global minimum. Furthermore, \cite{reweighted} notes that small values of $\epsilon$ (leading to a better approximation of the $\ell_0$ penalty) makes it more likely that the algorithm gets stuck at undesirable local minima.

\begin{figure}[t]
 \centering
 \subfigure[{Measurement Cutoffs}]{
  \includegraphics[width=0.45\textwidth]{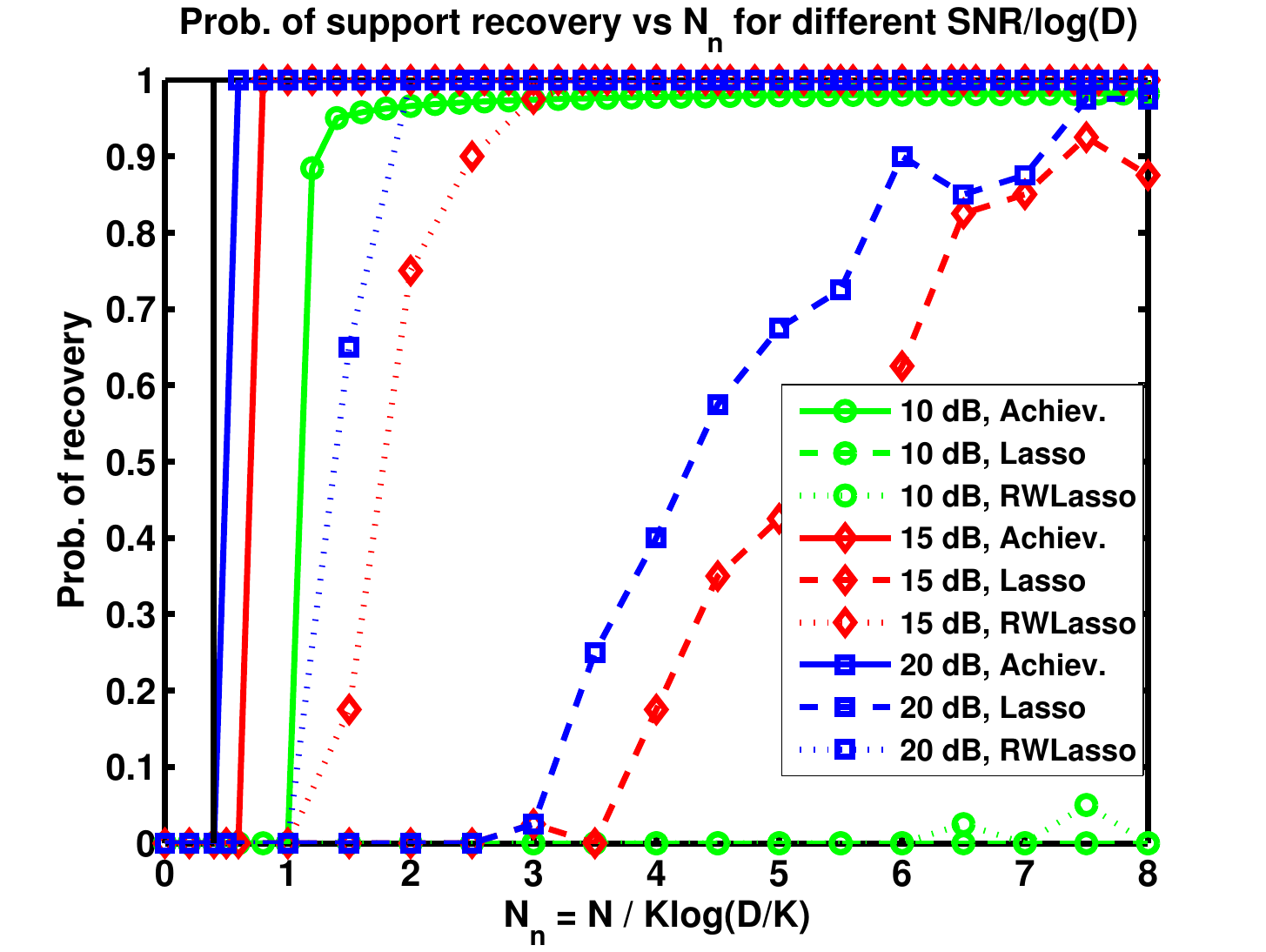}
   \label{fig:subfig11}
   }
 \subfigure[{Correlation Cutoffs}]{
  \includegraphics[width=0.45\textwidth]{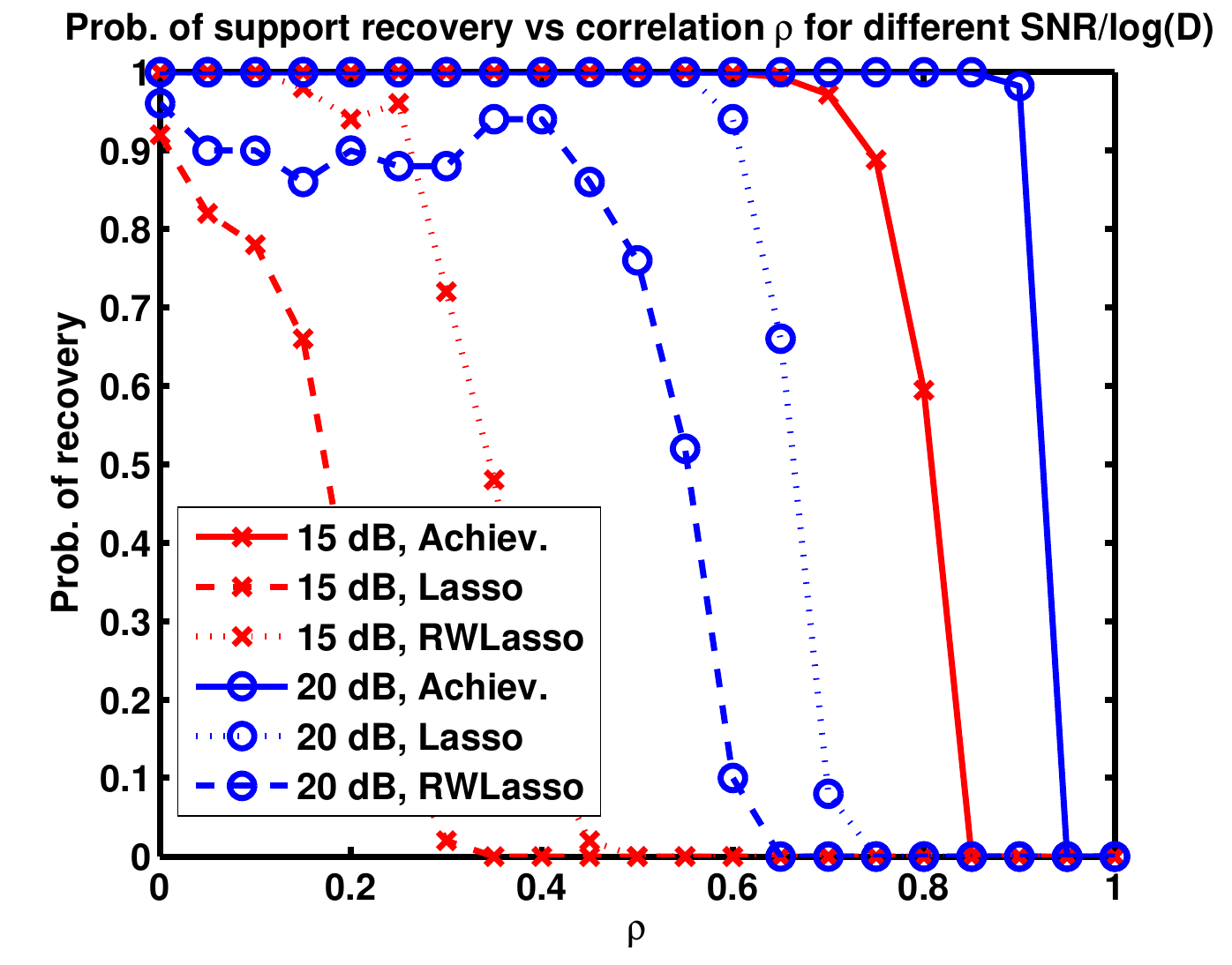}
   \label{fig:subfig12}
   }
 \label{fig:cs_bounds1}
 \caption{Comparison of information-theoretic bound vs.\ lasso and reweighted lasso.}
\end{figure}

We have chosen reweighted lasso for comparison with the information-theoretic bound since for CS, the optimal ML decoder can be equivalently written as an $\ell_0$ constrained least squares minimization for fixed $\bS$. Therefore we would expect a method like the reweighted lasso to better approach the achievable bound compared to lasso, as it aims to successively approximate the $\ell_0$ penalty while still being computationally efficient. We demonstrate that this is the case in our simulation results below. 

Figure \ref{fig:subfig11} plots the recovery bound for IID variables vs.\ lasso and reweighted simulation performance, for different number of measurements $N$. The probabilities of recovery for the lassos are computed over 40 iterations. Compared to lasso, our IT bound has a much sharper transition, while also being tighter, matching closely our lower bound (vertical line for $\SNR/\log D =$ 20 dB) obtained with Theorem \ref{thm:cs_I}. Interestingly, reweighted lasso nearly achieves our performance bounds for high SNR, however it fails in low SNR performance similar to lasso.
Note that the theoretical results in \cite{wainwright2, wainwright} for lasso are not strictly comparable since they require a significantly large SNR regime. Furthermore, the performance gap approaches infinity as we let $K$ approach $D$, implying lasso works strictly in sublinear regime.

Figure \ref{fig:subfig12} shows our probability of error bound vs.\ lasso performances for different values of the correlation coefficient $\rho$, where $N_n = 8$. The probabilities of recovery for lassos are computed over 50 iterations. This plot demonstrates clearly that while our bounds show tolerance to correlation up to a constant approaching $1$ (as seen from the sample complexity bound in Theorem \ref{thm:lcs}), lasso can tolerate at most $\rho=0.5$ correlation for exact recovery in this scenario, with very high SNR and $N$. Note that strongest results due to \cite{candesplan} require correlations to decay asymptotically to zero as $1/\log(D)$. 
Reweighted lasso shows better performance than lasso, however there is still a significant gap between the achievable correlation bound and the reweighted lasso performance, especially at 15 dB SNR.

\subsection{Support-OMP and noisy variables}

For the second set of experiments, we compare with a variant of the orthogonal matching pursuit (OMP) algorithm called support-OMP. This algorithm is proposed by \cite{caramanis} and shown to have good performance with theoretical guarantees for problems with noisy or missing observations of the sensing matrix $X^N$, as we consider in Section \ref{subsec:cs_noisy_bounds}.

\begin{figure}[t]
 \subfigure[Noise Variance Cutoffs]{
  \includegraphics[width=0.45\textwidth]{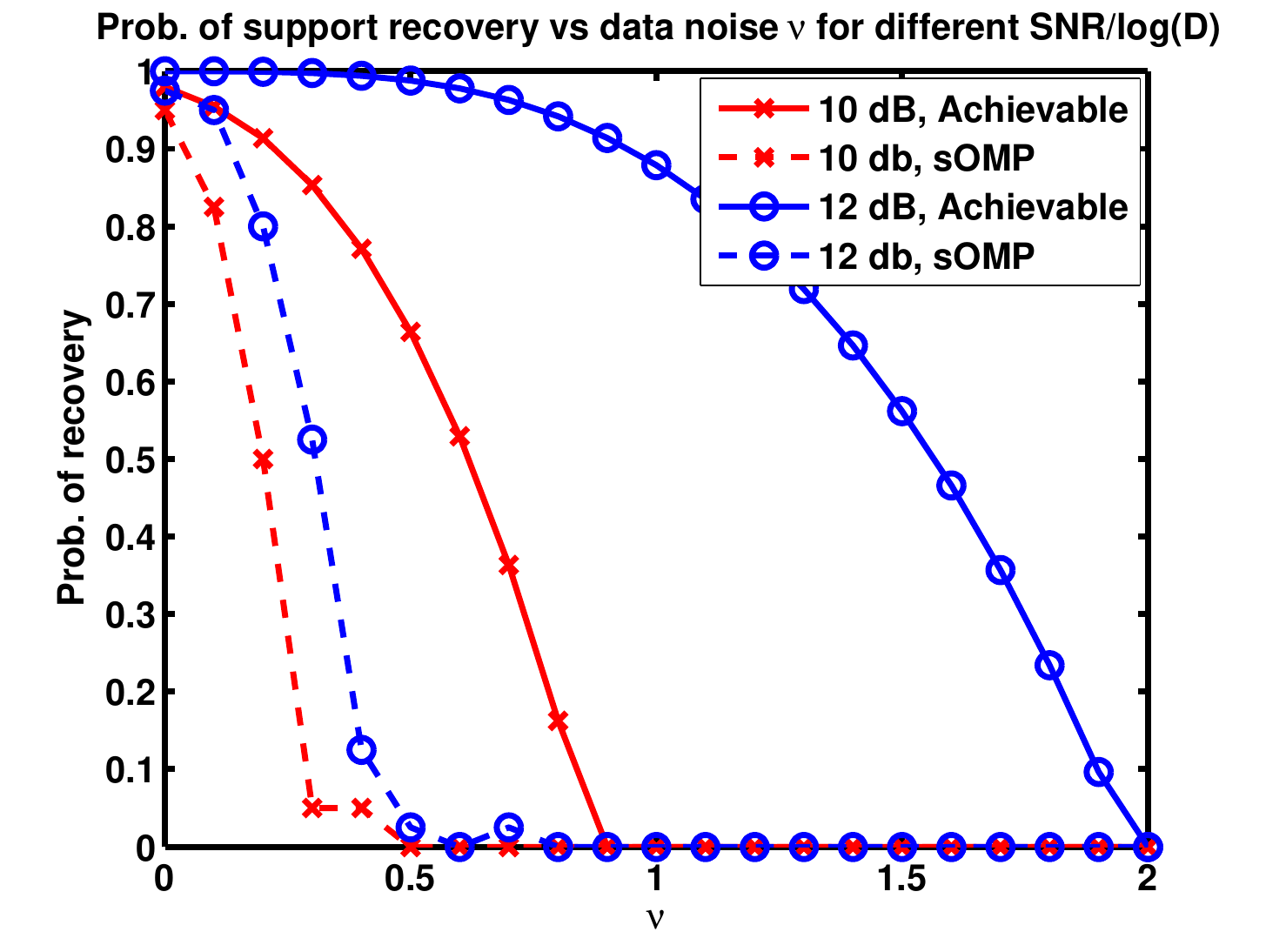}
   \label{fig:subfig21}
   }
 \subfigure[Effect of Both Noisy Data \& Correlation]{
  \includegraphics[width=0.45\textwidth]{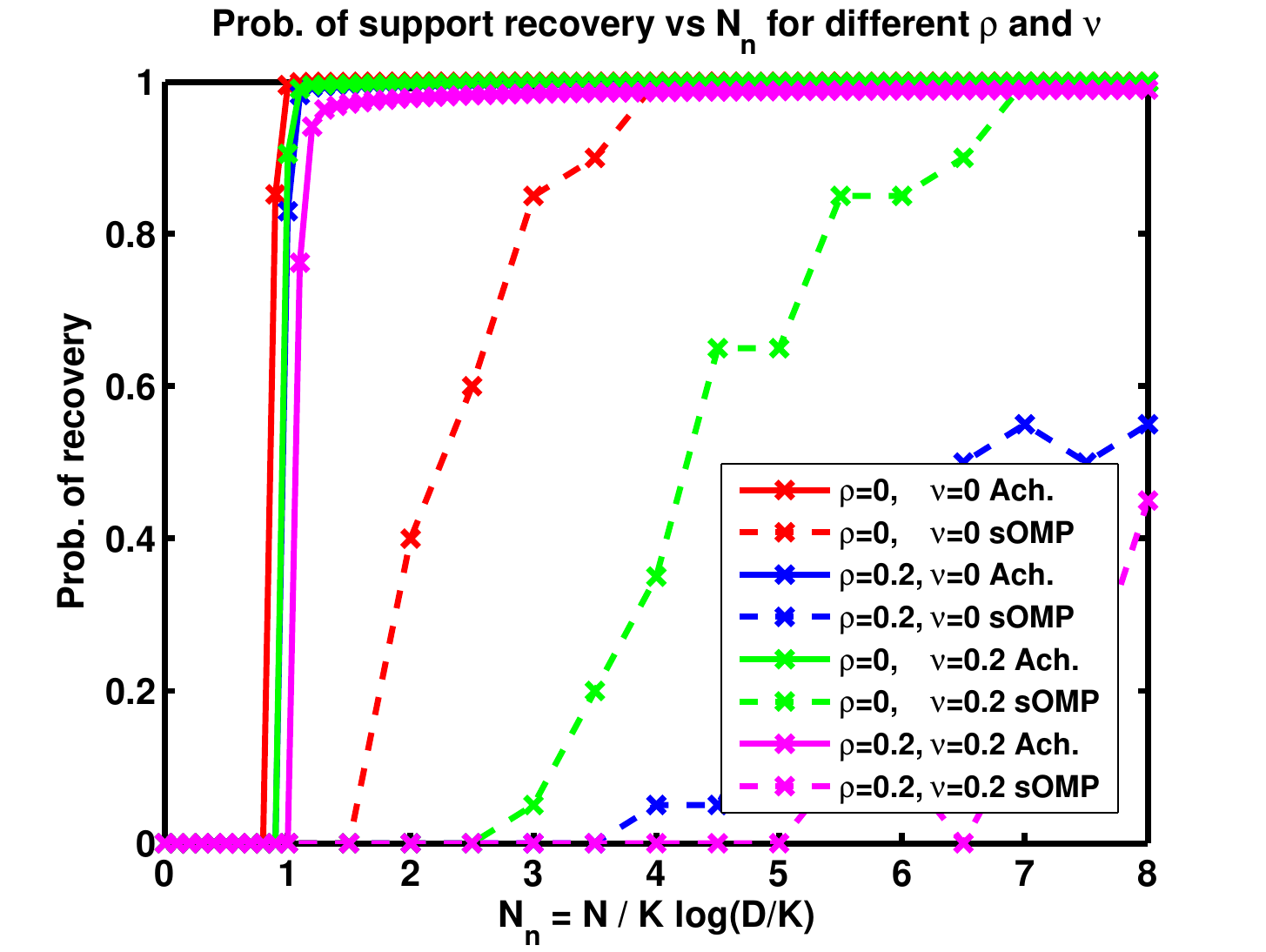}
   \label{fig:subfig22}
   }
 \label{fig:cs_bounds2}
 \caption{Comparison of information-theoretic bound vs.\ support-OMP.}\
\end{figure}

Figure \ref{fig:subfig21} shows the performance of support-OMP vs.\ information-theoretic bound, for noisy variables with different noise variances $\nu$. For support-OMP, the recovery probability is computed over 40 iterations. It can be seen that support-OMP performs reasonably well for noisy variables but fails in high variance noise, whereas our information-theoretic bound shows that recovery in much higher noise levels are achievable, especially with higher SNR.

A similar conclusion can be reached from Figure \ref{fig:subfig22}, where we plot recovery performance for both correlated and noisy variables. For support-OMP, the recovery probability is computed over 25 iterations. 
The gap is more pronounced for correlated variables compared to noisy variables, which shows support-OMP is highly affected by correlation and by variable noise to a lesser degree.

\section{Discussion}

We have presented a framework for analyzing sparse recovery problems that unifies linear and nonlinear observation models, dependent and non-Gaussian measurements matrices, and noisy data. This framework leads to a tight, exponential upper bound on the support recovery error probability and an explicit universal mutual information formula for computing the sample complexity of sparse recovery problems. The central theme here is ``inference of a discrete object (sparse support pattern) in a continuous world of observations.'' We unify sparse problems from an inference perspective by introducing a Markovian assumption. Our approach is not algorithmic and therefore must be used in conjunction with tractable algorithms. Nevertheless, it is useful for identifying gaps between existing algorithms and fundamental information. 

Although we consider sparse linear regression and its variants as applications in this paper, there are many other sparse recovery applications for which the framework we consider is applicable and the error bound or the sample complexity bounds we have described are explicitly computable through the formulas in Theorems \ref{thm:P(E)} and \ref{thm:main_theorem}. Some examples we have not included due to space considerations are group testing, quantized compressive sensing, multiple regression models or models with missing observations.

As we have shown our approach is also useful in understanding fundamental tradeoffs between different design parameters such as SNR, correlations, measurements matrices and noisy features. For instance, in the linear Gaussian setting we have shown that we could information theoretically tolerate up to constant correlation across different variables while existing results require vanishing correlation. The linear setting has also identified large sample complexity gaps between lasso, support-OMP and information theoretic bounds. Specifically, these gaps get larger as correlation and variable noise increases.

\newpage

\renewcommand{\theequation}{A.\arabic{equation}}
\section*{Appendix}
\setcounter{equation}{0}

We use lower-case $p(Y | X_S)$ notation for the conditional outcome distribution given the true subset of variables averaged over the latent variable $\beta_S$. In some cases when we would like to distinguish between the outcome distribution conditioned on different sets of variables we use $p_\omega(\,\cdot\, | \,\cdot\,)$ notation, to emphasize that the conditional distribution is conditioned on the given variables, assuming the true set $S$ is $S_\omega$. W.l.o.g.\ we assume the true set is $S_1$ for below proofs. Define ${\cal I} = \{1, \ldots, \binom{D}{K}\}$ as the collection of sets $\omega$ of size $K$.

\subsection*{Proof of Theorem \ref{thm:P(E)}}

First, note that $P(E) \leq \sum_{i=1}^K P(E_i)$, for $E$ and $E_i$ as defined. If we show separately for each $i$ and any $0 \leq \delta \leq 1$ that the following bound holds, then the theorem follows:

\begin{align}
P(E_i)\leq 2^{-N\left(E_o(\delta)-\delta\frac{\log\binom{D-K}{i}\binom{K}{i}}{N}\right)}.
\label{eq:error_exp}
\end{align}

Instead of the above bound, we prove a slightly weaker bound for expositional clarity, which is
\begin{align}
P(E_i)\leq 2^{-N\left(E_o(\delta)-\frac{\log\binom{D-K}{i}\binom{K}{i}}{N}\right)}.
\label{eq:error_exp_weak}
\end{align}
Note that the main difference between the above equation and the previous bound is the missing $\delta$ term multiplying the binomial expression. The main result follows along the same lines and we refer the reader to \cite{group_testing} for further details.

To prove this result we denote by ${\cal A}_i$ the set of indices corresponding to sets of $K$ variables that differ from the true set $S_1$ in exactly $i$ variables, i.e.,
\begin{align}
{\cal A}_i=\{\omega\in{\cal I}: |S_{1^c,\omega}|=i, |S_{\omega}|=K\}
\label{eq:setA}
\end{align}
We can establish that,
\begin{align} \label{eq.ratio}
\Pr[E_i | \omega_0 = 1, X^N_{S_1}, Y^N, \theta] &\leq
\sum_{\omega \in {\cal A}_i}\sum_{\substack{X^N_{S_{1^c,\omega}}}} P(X^N_{S_{1^c,\omega}}|\theta)\frac{p_{\omega}(Y^N|X^N_{S_{1,\omega}},X^N_{S_{1^c,\omega}})^s}{p_1(Y^N|X^N_{S_{1,\omega}},X^N_{S_{1,\omega^c}})^s}\\ \nonumber
& = \sum_{S_{1,\omega}} \sum_{S_{1^c,\omega}}\sum_{\substack{X^N_{S_{1^c,\omega}}}} P(X^N_{S_{1^c,\omega}}|\theta)\frac{p_{\omega}(Y^N|X^N_{S_{1,\omega}},X^N_{S_{1^c,\omega}})^s}{p_1(Y^N|X^N_{S_{1,\omega}},X^N_{S_{1,\omega^c}})^s}.
\end{align}
Inequality \eqref{eq.ratio} is established separately in the following section. It follows that, 
\begin{align}
\Pr[E_i | \omega_0 = 1, X^N_{S_1}, Y^N, \theta] & \leq \left (\sum_{S_{1,\omega}} \sum_{S_{1^c,\omega}}\sum_{\substack{X^N_{S_{1^c,\omega}}}} P(X^N_{S_{1^c,\omega}}|\theta)\frac{p_{\omega}(Y^N|X^N_{S_{1,\omega}},X^N_{S_{1^c,\omega}})^s}{p_1(Y^N|X^N_{S_{1,\omega}},X^N_{S_{1,\omega^c}})^s} \right )^{\delta} \label{l3.1:eq1} \\
& \leq \left (\sum_{S_{1,\omega}}\binom{D-K}{i} \sum_{\substack{X^N_{S_{1^c,\omega}}}} P(X^N_{S_{1^c,\omega}}|\theta)\frac{p_{\omega}(Y^N|X^N_{S_{1,\omega}},X^N_{S_{1^c,\omega}})^s}{p_1(Y^N|X^N_{S_{1,\omega}},X^N_{S_{1,\omega^c}})^s} \right )^{\delta} \label{l3.1:eq2} \\ 
& \leq \binom{D-K}{i}\sum_{S_{1,\omega}}\left (\sum_{\substack{X^N_{S_{1^c,\omega}}}} P(X^N_{S_{1^c,\omega}}|\theta)\frac{p_{\omega}(Y^N|X^N_{S_{1,\omega}},X^N_{S_{1^c,\omega}})^s}{p_1(Y^N|X^N_{S_{1,\omega}},X^N_{S_{1,\omega^c}})^s} \right )^{\delta}, ~ \forall s>0,\,\, 0\leq \delta \leq 1. \label{l3.1:eq3}
\end{align}
Inequality \eqref{l3.1:eq1} follows from the fact that $\Pr[E_i | \omega_0 = 1, X^N_{S_1}, Y^N, \theta] \leq 1$. Consequently, if $U$ is an upper bound of this probability then it follows that, $\Pr[E_i | \omega_0 = 1, X^N_{S_1}, Y^N, \theta] \leq U^{\delta}$ for $\delta \in [0,1]$. Inequality \eqref{l3.1:eq2} follows from symmetry, namely, the inner summation is only dependent on the values of $X^N_{S_{1^c,\omega}}$ and not on the items in the set $S_{1^c,\omega}$. There are exactly $\binom{D-K}{i}$ possible sets $S_{1^c,\omega}$ hence the binomial expression. Note that the sum over $S_{1,\omega}$ cannot be further simplified. This is due to the fact that $X^N_{S_{1,\omega}}$ is already specified since we have conditioned on $X^N_{S_1}$. Since $X^N_{S_1}$ is fixed, the inner sum need not be equal for all sets $S_{1,\omega},\omega\in{\cal A}_i$. Finally, \eqref{l3.1:eq3} follows from standard observation that sum of positive numbers raised to $\delta$-th power for $\delta < 1$ is smaller than the sum of the $\delta$-th power of each number.

We now substitute for the conditional error probability derived above and follow the steps below:
\begin{align} \nonumber
P(E_i) & = \int \sum_{X^N_{S_1}} \sum_{Y^N} P(\theta) P(X^N_{S_1}|\theta) p_1(Y^N|X^N_{S_1}) \Pr[E_i | \omega_0 = 1, X^N_{S_1}, Y^N, \theta] \dx \theta
\\ \nonumber
& \leq
\binom{D-K}{i} \int \sum_{S_{1,\omega}} \sum_{Y^N}\sum_{X^N_{S_1}} P(\theta) P(X^N_{S_1}|\theta) p_1(Y^N|X^N_{S_1}) \left (\sum_{\substack{X^N_{S_{1^c,\omega}}}} P(X^N_{S_{1^c,\omega}}|\theta)\frac{p_{\omega}(Y^N|X^N_{S_{1,\omega}},X^N_{S_{1^c,\omega}})^s}{p_1(Y^N|X^N_{S_{1,\omega}},X^N_{S_{1,\omega^c}})^s} \right )^{\delta} \dx \theta
\end{align}
Due to symmetry the summation over sets $S_{1,\omega}$ does not depend on $\omega$. Since there are $\binom{K}{K-i}$ sets $S_{1,\omega}$ we get,
\begin{align}
P(E_i)& \leq \nonumber
\binom{D-K}{i} \binom{K}{i} \int \sum_{Y^N}\sum_{X^N_{S_1}} P(\theta) P(X^N_{S_1}|\theta) p_1(Y^N|X^N_{S_1}) \left (\sum_{\substack{X^N_{S_{1^c,\omega}}}} P(X^N_{S_{1^c,\omega}}|\theta)\frac{p_{\omega}(Y^N|X^N_{S_{1,\omega}},X^N_{S_{1^c,\omega}})^s}{p_1(Y^N|X^N_{S_{1,\omega}},X^N_{S_{1,\omega^c}})^s} \right )^{\delta} \dx \theta
\\ \nonumber & \leq
\binom{D-K}{i} \binom{K}{i} \int \sum_{Y^N}\sum_{X^N_{S_{1,\omega^c}}}\sum_{X^N_{S_{1,\omega}}} P(\theta) P(X^N_{S_1}|\theta) p_1^{1-s\delta}(Y^N|X^N_{S_{1,\omega}},X^N_{S_{1,\omega^c}}) \\ & \nonumber ~~~~~~~~~~~~~~~~~~~~~~~~~~~~~~~~~~~~~~~~~~~~~~~~~~~~~~~~~~ \left( \sum_{\substack{X^N_{S_{1^c,\omega}}}} P(X^N_{S_{1^c,\omega}}|\theta) p_{\omega}(Y^N|X^N_{S_{1,\omega}},X^N_{S_{1^c,\omega}})^s \right)^{\delta} \dx \theta
\\ \nonumber & =
\binom{D-K}{i} \binom{K}{i} \int \sum_{Y^N} \sum_{X^N_{S_{1,\omega}}} P(\theta) P(X^N_{S_{1,\omega}}|\theta) \left ( \sum_{X^N_{S_{1,\omega^c}}} P(X^N_{S_{1,\omega^c}}|\theta) p_1^{1/(1+\delta)}(Y^N|X^N_{S_{1,\omega}},X^N_{S_{1,\omega^c}}) \right )^{1+\delta} \dx \theta
\end{align}
where the last step follows by letting $s = \frac{1}{1+\delta}$ and noting that from symmetry $X^N_{S_{1^c,\omega}}$ is just a dummy variable and can be replaced by $X^N_{S_{1,\omega^c}}$. This establishes the weaker bound in \eqref{eq:error_exp_weak}, by letting $\Sone = S_{1,\omega^c}$ and $\Stwo = S_{1,\omega}$. 

\qed

\subsubsection*{Proof of Equation \ref{eq.ratio}}

Let $\zeta_{\omega}$, $\omega\in{\cal A}_i$ denote the event where $\omega$ is more likely than $1$. Then, from the definition of ${\cal A}_i$, the $2$ encoded messages differ in $i$ variables. Hence
\begin{align}
\Pr[E_i | \omega_0 = 1, X^N_{S_1}, Y^N,\theta] & \leq P(\bigcup_{\substack{\omega\in{\cal A}_i}} \zeta_{\omega}) \leq \sum_{\substack{\omega\in{\cal A}_i}}P(\zeta_{\omega}) \nonumber
\end{align}
Now note that $X^N_{S_1}$ shares $(K-i)$ variables with $X^N_{S_{\omega}}$. Following the introduced notation, the common partition is denoted $X^N_{S_{1,\omega}}$, which is a $N \times (K-i)$ submatrix. The remaining $i$ rows which are in $X^N_{S_1}$ but not in $X^N_{S_{\omega}}$ are $X^N_{S_{1,\omega^c}}$. Similarly, $X^N_{S_{1^c,\omega}}$ corresponds to variables in $X^N_{S_{\omega}}$ but not in $X^N_{S_1}$. In other words $X^N_{S_1} = (X^N_{S_{1,\omega}},X^N_{S_{1,\omega^c}})$ and $X^N_{S_{\omega}} = (X^N_{S_{1,\omega}},X^N_{S_{1^c,\omega}})$, where the notation $(F^{N \times n_1};G^{N \times n_2})$ denotes an $N \times (n_1+n_2)$ matrix with a submatrix $F$ in the first $n_1$ columns and $G$ in the remaining $n_2$ columns. Thus,
\begin{align}
P(\zeta_{\omega}) & = \sum_{\substack{X^N_{S_{\omega}}:p(Y^N|X^N_{S_{\omega}})\geq p(Y^N|X^N_{S_1})}} P(X^N_{S_{\omega}}|X^N_{S_1},\theta)\nonumber\\
& \leq \sum_{\substack{X^N_{S_{1^c,\omega}}}} P(X^N_{S_{1^c,\omega}} | \theta)\frac{p(Y^N|X^N_{S_{\omega}})^s}{p(Y^N|X^N_{S_1})^s}~~~~ \forall s>0,~\forall\omega\in{\cal A}_i
\label{eq:zeta_m}
\end{align}
where by exchangeability, we have $P(X^N_{S_{\omega}}|X^N_\Sone, \theta) = P(X^N_{S_{1^c,\omega}}|X_\Sone^N, \theta) = P(X^N_{S_{1^c,\omega}}| \theta)$ and $\frac{p(Y^N|X^N_{S_{\omega}})^s}{p(Y^N|X^N_{S_1})^s} \geq 1$ for all $s > 0$, since $\frac{p(Y^N|X^N_{S_{\omega}})}{p(Y^N|X^N_{S_1})} \geq 1$. 

\qed

\subsection*{Proof of Theorem \ref{thm:main_theorem}}

We first derive the sufficiency bound, using the results of Theorem 2.1. 
To achieve that, we derive a sufficient condition for the error exponent of the error probability $P(E_i)$ in \eqref{eq:error_exp} to be positive and to drive the error probability to zero as $D\rightarrow\infty$. Specifically,
\begin{align}
Nf(\delta)= NE_o(\delta) - \delta\log\binom{D-K}{i}\binom{K}{i} \rightarrow\infty
\label{eq:f_rho_condition}
\end{align}
where
\[
f(\delta) = E_o(\delta)-\delta\frac{\log\binom{D-K}{i}\binom{K}{i}}{N}.
\]

To establish the sufficiency bound we follow the argument in \cite{gallager}. Note that $f(0)=0$. Since the function $f(\delta)$ is differentiable and has a power series expansion, for a sufficiently small $\delta$, we get by Taylor series expansion in the neighborhood $\delta = 0$ that,
\[
f(\delta) = f(0) + \delta\frac{df}{d\delta}\Big\vert_{\delta=0} + O(\delta^2)
\]

Note that
\begin{equation}
\frac{\partial E_o}{\partial \delta}\Big\vert_{\delta=0} = \frac{I(X_\Sone^N;Y^N | X_\Stwo^N,\theta)}{N}, \label{eq:Eo_I}
\end{equation}
which is shown in the next section.

We can further decompose $I(X_\Sone^N;Y^N | X_\Stwo^N,\theta)$ using the following chain of equalities:
\begin{align*}
I(X_\Sone^N ;  Y^N | X_\Stwo^N,\theta) + I(\beta_S ; X_\Sone^N | X_\Stwo^N, Y^N, \theta) & = I(X_\Sone^N ; Y^N, \beta_S | X_\Stwo^N, \theta) = I(X_\Sone^N ; \beta_S | \theta) + I(X_\Sone^N; Y^N | X_\Stwo^N, \beta_S, \theta) \\
& = N I(X_\Sone; Y | X_\Stwo, \beta_S, \theta),
\end{align*}
where the last equality is due to $X$ and $\beta_S$ being independent and $(X^N, Y^N)$ pairs being independent over $n$ given $\beta_S$. Therefore we have
\begin{equation} \label{eq:IT}
\frac{\partial E_o}{\partial \delta}\Big\vert_{\delta=0} = \frac{I(X_\Sone^N ; Y^N | X_\Stwo^N, \theta)}{N} = I(X_\Sone; Y | X_\Stwo, \beta_S, \theta) - \frac{I(\beta_S ; X_\Sone^N | X_\Stwo^N, Y^N, \theta)}{N}.
\end{equation}

Now assume that $N$ satisfies
\begin{equation}\label{eq:N_i}
N > (1+\epsilon) \frac{\log\binom{D-K}{i}\binom{K}{i}}{I(X_\Sone;Y | X_\Stwo,\beta_S, \theta)}.
\end{equation}
for any constant $\epsilon > 0$. 
We note that from the Lagrange form of the Taylor Series expansion (an application of the mean value theorem) we can write $E_o(\delta)$ in terms of its first derivative evaluated at zero and a remainder term, i.e.,
\begin{equation*}
E_o(\delta) = E_o(0) + \delta E_o'(0) + \frac{\delta^2}{2} E_o''(\psi)
\label{eq:lagrange}
\end{equation*}
for some $\psi \in [0,\delta]$. Hence, for the choice of $N$ in \eqref{eq:N_i} and using \eqref{eq:IT} we have
\begin{equation} \label{eq:N_f}
Nf(\delta)\geq N\left(\delta\frac{\epsilon}{1+\epsilon}I(X_\Sone; Y | X_\Stwo, \beta_S, \theta)-\delta^2 C I(X_\Sone;Y | X_\Stwo,\beta_S, \theta) - \delta \frac{I(\beta_S ; X_\Sone^N | X_\Stwo^N, Y^N, \theta)}{N} \right)
\end{equation}
where $C=\frac{|E_o''(\psi)|}{2I(X_\Sone; Y | X_\Stwo, \beta_S, \theta)}$ which might depend on $K$. 

A preliminary analysis of the necessary condition that we establish in the next section reveals that $N = \Omega(K \log D)$ is necessary, since $\log\binom{D-K+i}{i} = \Theta(i \log D)$ and $I(X_\Sone;Y | X_\Stwo,\beta_S, \theta) \leq H(Y) = O(1)$. Also, $I(\beta_S ; X_\Sone^N | X_\Stwo^N, Y^N, \theta) \leq H(\beta_S)$, which is constant with respect to $D$ since the observation model is only dependent on $K$ variables, due to the sparsity assumption of the observation model $P(Y|X)$. So we see that
\begin{equation*}
\frac{I(\beta_S ; X_\Sone^N | X_\Stwo^N, Y^N, \theta)}{N} = O\left(\frac{1}{\log D}\right)
\end{equation*}
which is always dominated by $I(X_\Sone; Y | X_\Stwo, \beta_S, \theta)$, which we assumed to be $\omega(1/\log D)$. Therefore we can rewrite \eqref{eq:N_f} as
\begin{equation*}
Nf(\delta)\geq N\left(\delta\left(\frac{\epsilon}{1+\epsilon} - o(1) \right) I(X_\Sone;Y |  X_\Stwo, \beta_S, \theta)-\delta^2 C I(X_\Sone;Y | X_\Stwo,\beta_S, \theta) \right).
\end{equation*}

Finally, if we choose $\delta\leq\frac{\epsilon'}{C}$, where $\epsilon' = \frac{\epsilon}{1+\epsilon}$, then $f(\delta)=\eta$ for some $\eta>0$ which does not depend on $D$ or $N$. It follows that $Nf(\delta)\rightarrow\infty$ as $D\rightarrow\infty$.

We have just shown that for fixed $K$, 
\[ N > (1+\epsilon)\cdot\frac{\log\binom{D-K}{i}\binom{K}{i}}{I(X_\Sone;Y | X_\Stwo,\beta_S, \theta)} \]
is sufficient to ensure an arbitrarily small $P(E_i)$. Now note that 
\[ (1+\epsilon) \binom{D-K+i}{i} \geq \binom{D-K}{i} \binom{K}{i} \]
asymptotically as $D\to \infty$ and $K$ is fixed, for any constant $\epsilon > 0$, which can be incorporated into the previous $\epsilon$ as both are arbitrary. 
Since the average error probability $P(E)\leq\sum_{i=1}^K P(E_i)$, it follows that if 
\[ N > (1+\epsilon) \max_{i=1,\ldots,K} \frac{\log\binom{D-K+i}{i}}{I(X_\Sone;Y | X_\Stwo,\beta_S, \theta)} \]
then for any fixed $K$, $\lim_{D\rightarrow\infty} P(E)=0$. Consequently, since this is true for any $K$, $\lim_{K\rightarrow\infty}\lim_{D\rightarrow\infty} P(E)=0$.

\qed

\vspace{10pt}
It is important to highlight the main difference between the analysis of the error probability for the problem considered herein and the channel coding problem. In contrast to channel coding, the codewords of a candidate set and the true set are not independent since the two sets could be overlapping. To overcome this difficulty, we separate the error events $E_i$, $i = 1,\ldots, K$, of misclassifying the true set in $i$ items. Then, for every $i$ we average over realizations of ensemble of codewords for every candidate set while holding fixed the partition common to these sets and the true set of variables.

\subsubsection*{Proof of Equation \ref{eq:Eo_I}}

We have
\begin{equation*}
E_o(\delta) = -\frac{1}{N} \log \sum_\theta \sum_{Y^N} \sum_{X_\Stwo^N} P(X_\Stwo^N | \theta) P(\theta) \left[\sum_{X_\Sone^N} P(X_\Sone^N | \theta) p(Y^N|X_\Sone^N,X_\Stwo^N)^{\frac{1}{1+\delta}}\right]^{1+\delta}, ~~~0\leq\delta\leq 1
\end{equation*}
where its derivative at $\delta = 0$ can be written as
\begin{align*}
\frac{\partial E_o}{\partial \delta}\Big\vert_{\delta=0} = -\frac{1}{N} & \frac{1}{\sum_{Y^N,X_\Sone^N,X_\Stwo^N,\theta} P(X_\Sone | \theta) P(X_\Stwo | \theta) P(\theta) p(Y^N | X_S^N)} \\
& \sum_{Y^N,X_\Stwo^N,\theta} P(X_\Stwo^N | \theta) P(\theta) \left( \sum_{X_\Sone^N} P(X_\Sone^N | \theta) p(Y^N|X_S^N) \left[ \log\left(\sum_{X_\Sone^N} P(X_\Sone^N | \theta) p(Y^N|X_S^N) \right) - \log p(Y^N|X_S^N)\right] \right).
\end{align*}

Noting that $P(X_\Sone^N | \theta) P(X_\Stwo^N | \theta) = P(X_\Sone^N, X_\Stwo^N | \theta)$ by the representation theorem and $p(Y^N|X_S^N) = p(Y^N|X_S^N,  \theta)$ by the independence of $Y$ and $\theta$ given $X_S$, above equality simplifies to
\begin{align*}
\frac{\partial E_o}{\partial \delta}\Big\vert_{\delta=0} & = -\frac{1}{N} \frac{1}{\sum_{Y^N,X_\Sone^N,X_\Stwo^N,\theta} P(Y^N, X_S^N, \theta)} \sum_{Y^N,X_\Sone^N, X_\Stwo^N,\theta} P(Y^N, X_S^N, \theta) \left( \log P(Y^N|X_\Stwo^N,\theta) - \log P(Y^N|X_S^N, \theta) \right) \\
& = \frac{1}{N} \sum_{Y^N,X_\Sone^N, X_\Stwo^N,\theta} P(Y^N, X_S^N, \theta) \log \frac{P(Y^N, X_\Stwo^N|X_\Sone^N, \theta)}{P(Y^N, X_\Stwo^N|\theta)} \\
& = \frac{I(X_\Sone^N ; X_\Stwo^N, Y^N | \theta)}{N} \\
& = \frac{I(X_\Sone^N ; Y^N | X_\Stwo^N, \theta)}{N},
\end{align*}
where the second equality follows by noting the first denominator is equal to $1$ and by adding and subtracting $\log P(X_\Stwo^N | \theta)$ inside the parenthesis. The third equality follows from the definition of mutual information. The final equality follows from the independence of $X_\Sone$ and $X_\Stwo$ given $\theta$.

\subsubsection*{Necessity bound}

The vector of outcomes $Y^N$ is probabilistically related to the index $\omega\in{\cal I}=\{1,2,\ldots,\binom{D}{K}\}$. Suppose $K-i$ elements of the salient set are revealed to us, denoted by $\Stwo$. 
From $X^N$ and $Y^N$ we estimate the set index $\omega$. Let the estimate be $\hat{\omega}=g(X^N,Y^N)$. Define the probability of error $P_e = P(E) = \Pr[\hat{\omega}\ne\omega]$.

$E$ is a binary random variable that takes the value $1$ in case of an error i.e., if $\hat{\omega}\ne\omega$, and $0$ otherwise, then using the chain rule of entropies \cite{coverbook} we have
\begin{align}
H(E,\omega | Y^N,X^N,\Stwo) & = H(\omega | Y^N,X^N,\Stwo) + H(E | \omega,Y^N,X^N,\Stwo)\nonumber\\
& = H(E | Y^N,X^N,\Stwo) + H(\omega|E,Y^N,X^N,\Stwo).
\label{eq:H(E,W|Y,X)}
\end{align}
The random variable $E$ is fully determined given $X^N$, $Y^N$, $\omega$ and $\Stwo$. It follows that $H(E | \omega,Y^N,X^N,\Stwo)=0$. Since $E$ is a binary random variable $H(E | Y^N,X^N,\Stwo)\leq 1$. Consequently, we can bound $H(\omega | E,Y^N,X^N,\Stwo)$ as follows,
\begin{align}
H(\omega | E,Y^N,X^N, \Stwo) & = P(E=0)H(\omega | E=0,Y^N,X^N,\Stwo) + P(E=1)H(\omega | E=1,Y^N,X^N,\Stwo)\nonumber\\
& \leq (1-P_e) \, 0 + P_e\log\left(\binom{D-K+i}{i}-1\right)\nonumber\\
& \leq P_e\log\binom{D-K+i}{i}.
\end{align}
The first inequality follows from the fact that revealing $K-i$ entries, and given that $E=1$, the conditional entropy can be upper bounded by the logarithm of the number of outcomes. From \eqref{eq:H(E,W|Y,X)}, we obtain the genie aided Fano's inequality
\begin{align}
H(\omega | Y^N,X^N,\Stwo) \leq 1 + P_e\log\binom{D-K+i}{i}
\label{eq:fano}
\end{align}

Note that for the left hand term, we have
\begin{align*}
H(\omega | Y^N,X^N,\Stwo) & = H(\omega | \Stwo) - I(\omega ; Y^N,X^N | \Stwo)\\
& = H(\omega | \Stwo) - I(\omega ; X^N | \Stwo) -  I(\omega ; Y^N | X^N, \Stwo) \\
& \stackrel{(a)}{=} H(\omega | \Stwo) -  I(\omega ; Y^N | X^N ,\Stwo) \\
&\stackrel{(b)}{=} H(\omega | \Stwo) -  (H(Y^N | X^N, \Stwo) -  H(Y^N | X^N, \omega))\\
&\stackrel{(c)}{=} H(\omega | \Stwo) -  (H(Y^N | X^N, \Stwo, \theta) -  H(Y^N | X^N, \omega, \theta))\\
&\stackrel{(d)}{\geq} H(\omega | \Stwo) -  (H(Y^N | X^N_{\Stwo}, \theta) -  H(Y^N | X^N_{S_\omega}, \theta))\\
&\stackrel{(e)}{=} H(\omega | \Stwo) -  I(X^N_{\Sone}; Y^N | X^N_{\Stwo}, \theta)\\
\end{align*}
where (a) follows from the fact that $X^N$ is independent of $\Stwo$ and $\omega$; (b) follows from the fact that conditioning with respect to $\omega$ includes conditioning with respect to $\Stwo$; (c) follows from the independence of $Y$ and $\theta$ given $X$; (d) follows from the fact that $Y^N$ depends on $\Stwo$ only through $X_{\Stwo}^N$ and similarly for the second term $Y^N$ depends on $\omega$ only through $X_\So^N$; the argument for (e) follows by definition.

From \eqref{eq:fano}, it then follows that
\begin{equation*}
H(\omega | \Stwo ) - I(X^N_\Sone ; Y^N | X^N_\Stwo, \theta) \leq 1+P_e\log\binom{D-K+i}{i}
\end{equation*}
and since the set ${\cal S}^2$ of $K-i$ variables is revealed, $\omega$ is uniformly distributed over the set of indices that correspond to sets of size $K$ containing ${\cal S}^2$. It follows that
\begin{equation*}
\log\binom{D-K+i}{i} - I(X^N_\Sone ; Y^N | X^N_\Stwo, \theta) \leq 1+P_e\log\binom{D-K+i}{i}.
\end{equation*}

Rewriting the above inequality, we have
\begin{equation}
P_e\geq 1-\frac{I(X^N_\Sone;Y^N|X^N_\Stwo, \theta)+1}{\log\binom{D-K+i}{i}}.
\end{equation}

Thus, for the probability of error to be asymptotically bounded away from zero, it is necessary that
\begin{align}
\log\binom{D-K+i}{i}\leq I(X^N_\Sone;Y^N|X^N_\Stwo, \theta) = N I(X_\Sone; Y | X_\Stwo, \beta_S, \theta) - I(\beta_S ; X_\Sone^N |  X_\Stwo^N, Y^N, \theta). \label{eq:LB_N}
\end{align}

Using \eqref{eq:IT}, we can see that
\begin{equation*}
N \geq \max_{i=1,\ldots,K} \frac{\log\binom{D-K+i}{i}}{I(X_\Sone; Y | X_\Stwo, \beta_S, \theta) - \frac{I(\beta_S ; X_\Sone^N |  X_\Stwo^N, Y^N, \theta)}{N} }
\end{equation*}
is a necessary condition for the number of samples $N$. Finally, since $I(\beta_S ; X_\Sone^N |  X_\Stwo^N, Y^N) \geq 0$, the following expression is a lower bound to the expression above, proving that it is a necessary condition for recovery, 
\[ N \geq \max_{i=1,\ldots,K} \frac{\log\binom{D-K+i}{i}}{I(X_\Sone; Y | X_\Stwo, \beta_S, \theta)}. \]

\qed

\subsection*{Continuous Variables}
\label{subsec:continuous}
Even though the results and proof ideas that were used in the above sections are fairly general, the proofs provided for sufficiency bounds were stated for discrete variables and outcomes. In this section we make the necessary generalizations to extend these proofs to continuous variable and observation models. We follow the methodology in \cite{gallager} and \cite{gallagerChapter}.

To simplify the exposition, we consider the extension to continuous variables in the special case of fixed and known $\beta_S$ and i.i.d.\ variables. Let $Q(X) = \prod_{i=1}^D Q(X_i)$ denote the joint distribution of variables $X$. The extensions to random $\beta_S$ and conditionally i.i.d.\ variables are straightforward. In this case, $I(X_\Sone;Y|X_\Stwo,\beta_S)$ reduces to $I(X_\Sone;Y|X_\Stwo)$ and $E_o(\delta)$ reduces to
\begin{equation}
E_o(\delta) = -\log\sum_{Y}\sum_{X_\Stwo}\left[\sum_{X_\Sone}Q(X_\Sone) p(Y,X_\Stwo|X_\Sone)^{\frac{1}{1+\delta}}\right]^{1+\delta} ~~~0\leq\delta\leq 1
\end{equation}
with $\frac{\partial E_o(\delta)}{\partial\delta}\Big\vert_{\delta=0} = I(X_\Sone;X_\Stwo,Y) = I(X_\Sone;Y|X_\Stwo)$, since $(X^{(n)}, Y^{(n)})$ pairs are independent across $n$ for fixed $\beta_S$.

Assume the continuous joint variable probability density $Q(X)$ with joint cumulative density function $F$ and the conditional probability density $p(Y=y|X_S=x)$ for the observation model, which is assumed to be a continuous function of both $x$ and $y$. 

Let $X^\prime \in {\cal X}^{\prime D}$ be the random vector and $Y^\prime \in {\cal Y}^\prime$ be the random variable generated by the quantization of $X \in {\cal X}^D = \mathbb{R}^D$ and $Y \in {\cal Y} = \mathbb{R}$ respectively, where each variable in $X$ is quantized to $L$ values and $Y$ quantized to $J$ values. Let $F^\prime$ be the joint cumulative density function of $X^\prime$. As before, let $\hat{S}(X^N,Y^N)$ be the ML decoder with continuous inputs with probability of making $i$ errors in decoding denoted by $P(E_i)$. Let $\hat{S}(X^{\prime N}, Y^{\prime N})$ be the ML decoder that quantizes inputs $X^N$ and $Y^N$ to $X^{\prime N}$ and $Y^{\prime N}$ and has a corresponding probability of error $P^\prime(E_i)$. Define
\begin{equation*}
E_o(\delta, X^\prime, Y^\prime) = -\log\sum_{y^\prime \in {\cal Y}^\prime}\sum_{x^{\prime}_\Stwo \in {\cal X}^{\prime K-i}}\left[\sum_{x^{\prime}_\Sone \in {\cal X}^{\prime i}}Q(x^{\prime}_\Sone) p(y^\prime,x^{\prime}_\Stwo|x^{\prime}_\Sone)^{\frac{1}{1+\delta}}\right]^{1+\delta},
\end{equation*}
\begin{equation*}
E_o(\delta, X, Y) = -\log \int_{\cal Y} \int_{{\cal X}^{K-i}} \left[ \int_{{\cal X}^i} Q(x_\Sone) p(y , x_\Stwo | x_\Sone)^{\frac{1}{1+\delta}} \dx x_\Sone \right]^{1+\delta} \dx x_\Stwo  \dx y,
\end{equation*}
where the indexing denotes the random variates which the error exponents are computed with respect to.

Utilizing the results in the proof of Theorem 3.1 for the discrete models, we will show the following for the continuous model 
\begin{equation}
\label{eq:cont_bound}
P(E_i)\leq 2^{-N\left(E_o(\delta, X, Y)-\delta\frac{\log\binom{D-K}{i}\binom{K}{i}}{N}\right)}.
\end{equation}
The rest of the proof will then follow as in the discrete case, by noting that $\frac{\partial E_o(\delta, X, Y)}{\partial\delta}\Big\vert_{\delta=0} = I(X_\Sone;X_\Stwo,Y)$, with the mutual information definition for continuous variables \cite{coverbook}.

Our strategy will be the following: we will increase the number of quantization levels for $Y^\prime$ and $X^\prime$ respectively and since discrete result \eqref{eq:error_exp} holds for any number of quantization levels, by taking limits we will be able to show that
\begin{equation}
P^\prime(E_i) \leq 2^{-N\left(E_o(\delta, X, Y)-\delta\frac{\log\binom{D-K}{i}\binom{K}{i}}{N}\right)}.
\end{equation}

Since $\hat{S}(X^N,Y^N)$ is the minimum probability of error decoder, any upper bound for $P^\prime(E_i)$ will also be an upper bound for $P(E_i)$, proving \eqref{eq:cont_bound}.

Assume $Y$ is quantized with the quantization boundaries denoted by $a_1, \ldots, a_{J-1}$, with $Y^\prime = a_j$ if $a_{j-1} < Y \leq a_j$. For convenience denote $a_0 = -\infty$ and $a_J = \infty$. Furthermore assume quantization boundaries are equally spaced, i.e. $a_j - a_{j-1} = \Delta_J$ for $2 \leq j \leq J-1$. Now we can write the following
\begin{alignat*}{2}
E_o(\delta, X^\prime, Y^\prime) = & -\log & & \sum_{j=1}^J \sum_{x^{\prime}_\Stwo} \left[ \sum_{x^{\prime}_\Sone}Q(x^{\prime}_\Sone) \left( \int_{a_{j-1}}^{a_j} p(y,x^{\prime}_\Stwo|x^{\prime}_\Sone) \dx y \right)^{\frac{1}{1+\delta}}\right]^{1+\delta} \\ 
= & -\log & & \Bigg\{ \sum_{j=2}^{J-1} \Delta_J \sum_{x^{\prime}_\Stwo} \left[ \sum_{x^{\prime}_\Sone}Q(x^{\prime}_\Sone) \left( \frac{\int_{a_{j-1}}^{a_j} p(y,x^{\prime}_\Stwo|x^{\prime}_\Sone) \dx y}{\Delta_J} \right)^{\frac{1}{1+\delta}}\right]^{1+\delta} \\
& & & \; + \sum_{x^{\prime}_\Stwo} \left[ \sum_{x^{\prime}_\Sone}Q(x^{\prime}_\Sone) \left( \int_{-\infty}^{a_1} p(y,x^{\prime}_\Stwo|x^{\prime}_\Sone) \dx y \right)^{\frac{1}{1+\delta}}\right]^{1+\delta} \\
& & & \; + \sum_{x^{\prime}_\Stwo} \left[ \sum_{x^{\prime}_\Sone}Q(x^{\prime}_\Sone) \left( \int_{a_{J-1}}^{\infty} p(y,x^{\prime}_\Stwo|x^{\prime}_\Sone) \dx y \right)^{\frac{1}{1+\delta}}\right]^{1+\delta} \Bigg\}.
\end{alignat*}

Let $J \to \infty$ and for each $J$ choose the sequence of quantization boundaries such that $\lim \Delta_J = 0$, $\lim a_{J-1} = \infty$, $\lim a_1 = -\infty$. Then the last two terms disappear and using the fundamental theorem of calculus, we obtain
\begin{equation}
\lim_{J \to \infty} E_o(\delta, X^\prime, Y^\prime) = E_o(\delta, X^\prime, Y) = -\log \int_{\cal Y} \sum_{x^{\prime}_\Stwo} \left[ \sum_{x^{\prime}_\Sone}Q(x^{\prime}_\Sone) p(y,x^{\prime}_\Stwo|x^{\prime}_\Sone)^{\frac{1}{1+\delta}}\right]^{1+\delta} \dx y.
\end{equation}

Although it is not necessary for our proof, it can also be shown that $E_o(\delta, X^\prime, Y^\prime)$ increases for finer quantizations of 
$Y^\prime$, therefore $E_o(\delta, X^\prime, Y)$ gives the smallest upper bound over $P^\prime(E_i)$ over the quantizations of $Y$.

We repeat the same procedure for $X$. Assume each variable $X_n$ in $X$ is quantized with the quantization boundaries denoted by $b_1, \ldots, b_{L-1}$, with $X_n^\prime = b_l$ if $b_{l-1} < X_n \leq b_l$. For convenience denote $b_0 = -\infty$ and $b_L = \infty$. Furthermore assume quantization boundaries are equally spaced, i.e. $b_l - b_{l-1} = \Delta_L$ for $2 \leq l \leq L-1$. Then we can write
\begin{alignat}{2}
E_o(\delta, X^\prime, Y) = & -\log \int_Y && \sum_{l=1}^L \left[ \sum_{x^{\prime}_\Sone}Q(x^{\prime}_\Sone) \left( \int_{b_{l-1}}^{b_l} p(y,x_\Stwo|x^{\prime}_\Sone) \dx x_\Stwo \right)^{\frac{1}{1+\delta}} \right]^{1+\delta} \dx y \nonumber \\
= & -\log \int_{\cal Y} && \sum_{l=1}^L \left[ \int_{{\cal X}^i} \left( \int_{b_{l-1}}^{b_l} p(y,x_\Stwo|x_\Sone) \dx x_\Stwo \right)^{\frac{1}{1+\delta}} \dx F^\prime(x_\Sone) \right]^{1+\delta} \dx y \label{eq:cont_xs1} \\
= & -\log \int_{\cal Y} && \Bigg\{ \sum_{l=2}^{L-1} \Delta_L \left[ \int_{{\cal X}^i} \left( \frac{\int_{b_{l-1}}^{b_l} p(y,x_\Stwo|x_\Sone) \dx x_\Stwo}{\Delta_L} \right)^{\frac{1}{1+\delta}} \dx F^\prime(x_\Sone) \right]^{1+\delta} \nonumber \\
&&& \; + \int_{{\cal X}^i} \left( \int_{-\infty}^{b_1} p(y,x_\Stwo|x_\Sone) \dx x_\Stwo \right)^{\frac{1}{1+\delta}} \dx F^\prime(x_\Sone) \nonumber \\
&&& \; + \int_{{\cal X}^i} \left( \int_{b_{L-1}}^{\infty} p(y,x_\Stwo|x_\Sone) \dx x_\Stwo \right)^{\frac{1}{1+\delta}} \dx F^\prime(x_\Sone)  \Bigg\} \dx y. \nonumber 
\end{alignat}
where \eqref{eq:cont_xs1} follows with $F^\prime(x_\Sone)$ being the step function which represents the cumulative density function of the quantized variables $X^{\prime}_\Sone$.

Let $L \to \infty$, for each $L$ choose a set of quantization point such that $\lim \Delta_L = 0$, $\lim b_{L-1} = \infty$, $\lim b_1 = -\infty$. Again, the second and third terms disappear and the first sum converges to the integral over $X_\Stwo$. Note that $p(y,x_\Stwo|x_\Sone)$ is a continuous function of all its variables since it was assumed that $Q(x)$ and $p(y|x)$ were continuous. Also note that $\lim_{L \to \infty} F^\prime = F$, which implies the weak convergence of the probability measure of $X^\prime$ to the probability measure of $X$. Given these facts, using the portmanteau theorem we obtain that $E_{F^\prime} \left[ p(Y,X_\Stwo|X_\Sone)\right] \to E_F\left[ p(Y,X_\Stwo|X_\Sone) \right]$, which leads to
\begin{equation}
\lim_{L \to \infty} E_o(\delta, X^\prime, Y) = -\log \int_{\cal Y} \int_{{\cal X}^{K-i}} \left[ \int_{{\cal X}^i} p(y , x_\Stwo | x_\Sone)^{\frac{1}{1+\delta}} \dx F(x_\Sone) \right]^{1+\delta} \dx x_\Stwo \dx y = E_o(\delta, X, Y).
\end{equation}

This leads to the following result, completing the proof.
\begin{equation}
P(E_i) \leq P^\prime(E_i) \leq \lim_{J, L \to \infty} 2^{-N\left( E_o(\delta, X^\prime, Y^\prime) - \delta \frac{\log\binom{D-K}{i}\binom{K}{i}}{N} \right)} = 2^{-N\left( E_o(\delta, X, Y) - \delta \frac{\log\binom{D-K}{i}\binom{K}{i}}{N} \right)}.
\end{equation}

\subsection*{Proof of Theorem \ref{thm:cs_bound}}

To derive the upper bound on error probability, we compute $E_o(\delta)$ explicitly and replace it in Theorem 2.1. First we compute for the easier case, with fixed $\beta_S = \sigma$. In this case, note that $(X, Y)$ pairs are independent across samples and
\begin{align*}
E_o(\delta) = - \log \int_\theta P(\theta) \int_{Y} \int_{X_\Stwo} P(X_\Stwo | \theta) \left[\int_{X_\Sone} P(X_\Sone | \theta) p(Y|X_\Sone,X_\Stwo)^{\frac{1}{1+\delta}}\dx X_\Sone \right]^{1+\delta} \dx X_\Stwo \dx Y \dx \theta, ~~~0\leq\delta\leq 1.
\end{align*}

For the correlated Gaussian variables, this reduces to
\begin{align*} E_o(\delta) = - \log & \int_\mu {\cal N}(\mu; 0, \rho / N) \int_{Y} \int_{X_\Stwo} {\cal N}(x_2; (K-i)\mu, (K-i)(1-\rho)/N )  \\
& \left[\int_{X_\Sone} {\cal N}(x_1; i\mu, i(1-\rho)/N ) {\cal N}(y - \sigma(x_1+x_2); 0, 1/\SNR )^{\frac{1}{1+\delta}} \dx x_1\right]^{1+\delta} \dx x_2 \dx y \dx \mu.
\end{align*}

As the first step, we input the Gaussian distributions and take the integral inside the brackets over $x_1$, which gives us
\begin{align*}
\Big[ \int_{X_\Sone}  {\cal N}(x_1; i\mu, i(1-\rho)/N ) & {\cal N}(y - \sigma(x_1+x_2); 0, 1/\SNR )^{\frac{1}{1+\delta}} \dx x_1 \Big]^{1+\delta} \\
& = \frac{\left( \sqrt{\frac{1}{\SNR\sigma^2}} \right)^\delta}{ \sigma \sqrt{2 \pi} \left( \sqrt{\frac{i(1-\rho)}{N(1+\delta)} + \frac{1}{\SNR\sigma^2}} \right)^{1+\delta}} \exp \left(- \frac{(\frac{y}{\sigma} - x_2 - i \mu)^2}{2\left( \frac{i(1-\rho)}{N(1+\delta)} + \frac{1}{\SNR\sigma^2}\right) } \right).
\end{align*}

By plugging in this expression and integrating over $x_2$, we then have
\begin{align*}
\int_{X_\Stwo} & {\cal N}(x_2; (K-i)\mu, (K-i)(1-\rho)/N ) \Big[ \int_{X_\Sone}  {\cal N}(x_1; i\mu, i(1-\rho)/N ) {\cal N}(y - \sigma(x_1+x_2); 0, 1/\SNR )^{\frac{1}{1+\delta}} \dx x_1 \Big]^{1+\delta} \dx x_2 \\
& = \frac{1}{\sigma \sqrt{2 \pi}} \left( \frac{1}{\sqrt{1 + \frac{i(1-\rho)\SNR\sigma^2}{N(1+\delta)}}} \right)^\delta \frac{1}{\sqrt{\frac{(K-i)(1-\rho)}{N} + \frac{i(1-\rho)}{N(1+\delta)} + \frac{1}{\SNR\sigma^2}}} \exp \left( - \frac{(y - \sigma K \mu)^2}{2 \sigma^2 \left(\frac{(K-i)(1-\rho)}{N} + \frac{i(1-\rho)}{N(1+\delta)} + \frac{1}{\SNR\sigma^2} \right) }\right)
\end{align*}

Integrating the above expression over $y$, we are left with
\[ \left( \frac{1}{\sqrt{1 + \frac{i(1-\rho)\SNR\sigma^2}{N(1+\delta)}}} \right)^\delta, \]
which no longer depends on $\mu$, therefore the expectation over $\mu$ is equal to the above expression and finally we have
\[ E_o(\delta) = \frac{\delta}{2} \log \left( 1 + (1-\rho) \frac{i \sigma^2 \SNR}{N (1+\delta)}\right), \]
for any $0 \leq \delta \leq 1$.

Now we will show a lower bound on the error exponent $E_o(\delta)$ for the case where $\beta_S$ is random and IID ${\cal N}(0, \sigma^2)$. In this case, $Y^{(n)}$ are not independent across $n$. In order to lower bound $E_o$, we first upper bound the observation probability such that,
\[ p(Y^N | X_\Sone^N , X_\Stwo^N)^{\frac{1}{1+\delta}} = \left( \int_{\bS} P(\bS) P(Y^N | X_\Sone^N, X_\Stwo^N, \bS) \dx \bS \right)^{\frac{1}{1+\delta}} \leq \int_{\bS} P(\bS)^{\frac{1}{1+\delta}} P(Y^N | X_\Sone^N, X_\Stwo^N, \bS)^{\frac{1}{1+\delta}} \dx \bS  \]
which follows from the subadditivity of exponent ${\frac{1}{1+\delta}}$. A lower bound on $E_o$ is then given by
\begin{align*} 
E_o(\delta) \geq -\frac{1}{N} \log M^{1+\delta} \int_{\theta^N} P(\theta^N) & \int_{Y^N} \int_{X_\Stwo^N} P(X_\Stwo^N | \theta^N) \\
& \left[\int_\bS \frac{P(\bS)^{\frac{1}{1+\delta}}}{M} \int_{X_\Sone^N} P(X_\Sone^N | \theta^N) P(Y^N|X_\Sone^N,X_\Stwo^N, \bS)^{\frac{1}{1+\delta}} \dx X_\Sone^N \dx \bS \right]^{1+\delta} \dx X_\Stwo^N \dx Y^N \dx \theta 
\end{align*}
where $M = \int P(\bS)^{\frac{1}{1+\delta}} \dx \bS$ and then by Jensen's inequality, it follows that
\begin{align} 
E_o(\delta) \geq -\frac{1}{N} \log M^\delta \int_\bS P(\bS)^{\frac{1}{1+\delta}} \Bigg( &  \int_\theta P(\theta)\int_{Y} \int_{X_\Stwo} P(X_\Stwo | \theta) \nonumber \\
& \left[\int_{X_\Sone} P(X_\Sone | \theta) P(Y|X_\Sone,X_\Stwo, \bS)^{\frac{1}{1+\delta}} \dx X_\Sone \right]^{1+\delta} \dx X_\Stwo \dx Y \dx \theta \Bigg)^N \dx \bS \label{eq:Eo_lb}
\end{align}
where we also used the independence of $(X^{(n)}, Y^{(n)})$ across $n$ given $\bS$.

We start by taking the integral inside the square brackets. For the linear model set-up we have,
\begin{align*}
\int_{X_\Sone} & P(X_\Sone | \theta) P(Y|X_\Sone,X_\Stwo, \bS)^{\frac{1}{1+\delta}} \dx X_\Sone = \int_{\mathbb{R}^i} {\cal N}\left( x; \mu 1_i, \frac{1-\rho}{N} I_i \right) {\cal N}\left(y - x^\top \bone - x_2^\top \btwo; 0, 1/\SNR \right)^{\frac{1}{1+\delta}} \dx x \\
& = \left( \frac{1}{\sqrt{2\pi A}}\right)^i \left( \frac{\sqrt{\SNR}}{\sqrt{2\pi}} \right)^{\frac{1}{1+\delta}} \int_{\mathbb{R}^i} \exp \left( - \frac{(x-\mu 1_i)^\top(x-\mu 1_i)}{2 A} \right) \exp \left( - \frac{(y - x^\top \bone - x_2^\top \btwo)^2}{2B} \right) \dx x \\
& = \left( \frac{1}{\sqrt{2\pi A}}\right)^i \left( \frac{\sqrt{\SNR}}{\sqrt{2\pi}} \right)^{\frac{1}{1+\delta}} \int_{\mathbb{R}^i} \exp \left( - \frac{x^\top x}{2 A}  - \frac{(x^\top \bone + C)^2}{2B} \right) \dx x\\
& = \left( \frac{1}{\sqrt{2\pi A}}\right)^i \left( \frac{\sqrt{\SNR}}{\sqrt{2\pi}} \right)^{\frac{1}{1+\delta}} \int_{\mathbb{R}^i} \exp \left( - \frac{1}{2}(x + (BD)^{-1} AC \bone)^\top \frac{D}{A} (x + (BD)^{-1} AC \bone) \right) \exp \left( - \frac{C^2}{2E} \right) \dx x
\end{align*}
where $A = \frac{1-\rho}{N}$, $B = \frac{1+\delta}{\SNR}$, $C = x_2^\top\btwo + \mu 1_i^\top \bone - y$, $D = I_i + \frac{A}{B} \bone \bone^\top$ and $E = \frac{B}{1 - \frac{A}{B} \bone^\top D^{-1} \bone}$. Then taking the integral, some terms on the left cancel and we have
\begin{equation} \label{eq:cs_pyx}
\int_{X_\Sone} P(X_\Sone | \theta) P(Y|X_\Sone,X_\Stwo, \bS)^{\frac{1}{1+\delta}} \dx X_\Sone = \left( \frac{\sqrt{\SNR}}{\sqrt{2\pi}} \right)^{\frac{1}{1+\delta}} \frac{1}{\sqrt{|D|}} \exp \left( - \frac{C^2}{2E} \right).
\end{equation}

Writing the second integral that is over $X_\Stwo = x_2$, we then have
\begin{align*}
\int_{X_\Stwo} & P(X_\Stwo | \theta) \left[ \int_{X_\Sone} P(X_\Sone | \theta) P(Y|X_\Sone,X_\Stwo, \bS)^{\frac{1}{1+\delta}} \dx X_\Sone \right]^{1+\delta} \dx X_\Stwo \\  
& = \sqrt{\frac{\SNR}{2\pi}} \frac{1}{\sqrt{|D|}^{(1+\delta)}} \int_{\mathbb{R}^{K-i}} {\cal N}(x ; \mu 1_{K-i}, A I_{K-i}) \exp \left( -\frac{(x^\top \btwo + \mu 1_i^\top \bone - y)^2}{2E^\prime} \right) \dx x \\
& = \sqrt{\frac{\SNR}{2\pi}} \frac{1}{\sqrt{|D|}^{(1+\delta)}} \left(\frac{1}{\sqrt{2\pi A}}\right)^{K-i} \int_{\mathbb{R}^{K-i}} \exp\left( -\frac{x^\top x}{2A} - \frac{(x^\top \btwo + F)^2}{2E^\prime}\right) \dx x \\
& = \sqrt{\frac{\SNR}{2\pi}} \frac{1}{\sqrt{|D|}^{(1+\delta)}} \left(\frac{1}{\sqrt{2\pi A}}\right)^{K-i} \int_{\mathbb{R}^{K-i}} \exp\left(-\frac{1}{2}(x + (E^\prime G)^{-1}AF \btwo)^\top \frac{G}{A} (x + (E^\prime G)^{-1}AF \btwo)\right) \exp \left( -\frac{F^2}{2H} \right) \dx x
\end{align*}
where $E^\prime = \frac{E}{1+\delta}$, $F = \mu 1_K^\top \bS - y$, $G = 1 + \frac{A}{E^\prime} \btwo \btwo^\top$ and $H = \frac{E^\prime}{1 - \frac{A}{E^\prime} \btwo^\top G^{-1} \btwo}$. Again, evaluating the integral, we obtain
\[ \int_{X_\Stwo} P(X_\Stwo | \theta) \left[ \int_{X_\Sone} P(X_\Sone | \theta) P(Y|X_\Sone,X_\Stwo, \bS)^{\frac{1}{1+\delta}} \dx X_\Sone \right]^{1+\delta} \dx X_\Stwo = \sqrt{\frac{\SNR}{2\pi}} \frac{1}{\sqrt{|D|}^{(1+\delta)}} \frac{1}{\sqrt{|G|}} \exp\left(-\frac{F^2}{2H}\right). \]

Integrating the above expression w.r.t.\ $Y = y$, we see that the result is independent of $\theta = \mu$, and therefore
\[ \int_\theta P(\theta) \int_{Y} \int_{X_\Stwo} P(X_\Stwo | \theta) \left[\int_{X_\Sone} P(X_\Sone | \theta) P(Y|X_\Sone,X_\Stwo, \bS)^{\frac{1}{1+\delta}} \dx X_\Sone \right]^{1+\delta} \dx X_\Stwo \dx Y \dx \theta = \frac{\sqrt{\SNR}}{\sqrt{|D|}^{(1+\delta)}} \sqrt{\frac{H}{|G|}}. \] 

By the matrix determinant lemma, we have $|D| = 1 + \frac{A}{B} \bone^\top \bone$ and by the Sherman-Morrison formula, $D^{-1} = I_i - \frac{\bone \bone^\top}{\frac{B}{A} + \bone^\top \bone}$. Similarly, $|G| = 1 + \frac{A}{E^\prime} \btwo^\top \btwo$ and $G^{-1} = I_i - \frac{\btwo \btwo^\top}{\frac{E^\prime}{A} + \btwo^\top \btwo}$. By plugging in these expressions, we can then see that $E^\prime = \frac{B |D|}{1+\delta}$ and $H = E^\prime |G|$. We simplify the above expression to obtain
\begin{equation} \label{eq:cs_beta}
\frac{\sqrt{\SNR}}{\sqrt{|D|}^{(1+\delta)}} \sqrt{\frac{H}{|G|}} = \frac{\sqrt{\SNR}}{\sqrt{|D|}^{(1+\delta)}} \sqrt{\frac{B |D|}{1+\delta}} = \left( \frac{1}{\sqrt{|D|}} \right)^\delta = \left( 1 + (1-\rho) \frac{\SNR \bone^\top \bone}{N(1+\delta)}\right)^{-\delta/2}.
\end{equation}

Note that this expression is analogous to the bound we obtained for the fixed case, since $E[\bone^\top \bone] = i \sigma^2$. With the above bound, we will now show a lower bound on $E_o(\delta)$ for $\delta=1$ and $\sigma^2 = \frac{1}{8\pi}$. We note that we choose this $\sigma^2$ without loss of generality, since for any value or scaling of $\sigma$ can be incorporated into the $\SNR$ of the problem to obtain an equivalent model, such that $\SNR\sigma^2$ is fixed. This result can also be shown without the assumption on $\sigma^2$, but the specific bounding methods we use utilize this assumption. To this effect, we analyze the equivalent problem with parameters $\SNR^\prime = \SNR \sigma^2 8\pi$ and $\sigma^{\prime 2} = \frac{1}{8\pi}$. 
Note that with this choice of $\sigma^{\prime 2}$ and $\delta$, $M = \int_{\mathbb{R}^K} P(\bS)^{\frac{1}{2}} \dx \bS = 1^K = 1$. Using \eqref{eq:Eo_lb}, we now write,
\begin{align*}
E_o(1) & \geq -\frac{1}{N} \log M \int_{\mathbb{R}^K} P(\bS)^{\frac{1}{2}} \left( 1 + (1-\rho) \frac{\SNR \bone^\top \bone}{2N}\right)^{-\frac{N}{2}} \dx \bS \\
& = -\frac{1}{N} \log \int_{\mathbb{R}^i} P(\bone)^{\frac{1}{2}} \left( 1 + (1-\rho) \frac{\SNR^\prime \bone^\top \bone}{2N}\right)^{-\frac{N}{2}} \dx \bone \\
& = -\frac{1}{N} \log (\sqrt{4})^\frac{i}{2} \int_{\mathbb{R}^i} \exp \left[ - \frac{\bone^\top \bone}{4\sigma^{\prime 2}} \right]\left( 1 + (1-\rho)\frac{\SNR^\prime \bone^\top \bone}{2N} \right)^{-\frac{N}{2}} \dx \bone \\
& \geq -\frac{1}{N} \log(\sqrt{4})^\frac{i}{2} \left( \sqrt{8\pi \sigma^{\prime 2}} \right)^\frac{iN}{2} \int_{\mathbb{R}^i} \left[\left( \frac{1}{\sqrt{8\pi \sigma^{\prime 2}}} \right)^i \exp \left[ - \frac{\bone^\top \bone}{8\sigma^{\prime 2}}\right]\left( 1 + (1-\rho)\frac{\SNR^\prime \bone^\top \bone}{2N} \right) \right]^{-\frac{N}{2}} \dx \bone \\
& \geq -\frac{1}{N} \log 4^{\frac{i}{4}} \left[ \int_{\mathbb{R}^i} \left( \frac{1}{\sqrt{8\pi \sigma^{\prime 2}}} \right)^i \exp \left[ -\frac{\bone^\top \bone}{8\sigma^{\prime 2}} \right] \left( 1 + (1-\rho)\frac{\SNR^\prime \bone^\top \bone}{2N} \right)\dx \bone \right]^{-\frac{N}{2}} \\
& = \frac{1}{2} \log \left( 1 + (1-\rho)\frac{2i \SNR \sigma^2}{N} \right) - \frac{i}{4N} \log 4.
\end{align*}
The first equality follows by taking $\btwo$ out of the integral and noting that $\int_{\mathbb{R}^{K-i}}^{} P(\btwo)^{\frac{1}{2}} \dx \btwo = 1^{K-i} = 1$. We obtain the second equality by expanding $P(\bone)^{\frac{1}{2}}$. We upper bound $\exp \left[ - \frac{\bone^\top \bone}{8\sigma^{\prime 2}}\right]$ by $\exp \left[ - \frac{\bone^\top \bone}{8\sigma^{\prime 2}}\right]^{-\frac{N}{2}}$ where $0 \leq \exp \left[ - \frac{\bone^\top \bone}{8\sigma^{\prime 2}}\right] \leq 1$ to obtain the first inequality and the second one follows by the superadditivity of exponentiating with $-\frac{N}{2}$. Finally, we note that the integral is an expectation w.r.t.\ $\bone \sim {\cal N}(0, 4 \sigma^{\prime 2} I_i)$ and obtain the last equality, where we also replace $\SNR^\prime$ and $\sigma^{\prime 2}$.

\qed

\subsection*{Proof of Lemma \ref{thm:cs_I}}

Note the following equalities,
\begin{align*}
I(X_\Sone ; Y | X_\Stwo, \beta_S, \mu) & = h(Y | X_\Stwo, \beta_S, \mu) - h(Y | X_S, \beta_S, \mu) \\
& = h\left( X_\Sone^\top \beta_\Sone + W | \beta_\Sone, \mu \right) - h(W) \\
& = E_{\beta_\Sone, \mu} \left[ \frac{1}{2} \ln\left( 2\pi e \left(\text{var}\left(X_\Sone^\top \beta_\Sone | \beta_\Sone, \mu\right) \, \beta_\Sone^\top \beta_\Sone + \frac{1}{\SNR}\right)\right) \right] - \frac{1}{2} \ln \left( 2\pi e \; \frac{1}{\SNR}\right) \\
& = E_{\beta_\Sone} \left[ \frac{1}{2} \ln \left( 1 + (1-\rho)\frac{\beta_\Sone^\top \beta_\Sone \SNR}{N} \right) \right],
\end{align*}
where the second equality follows from the independence of $X_\Sone$ and $X_\Stwo$ given $\mu$ and the last equality follows from the fact that $\text{var}(X_\Sone^\top \beta_\Sone | \beta_\Sone, \mu) = \beta_\Sone^\top E[U_\Sone U_\Sone^\top]\beta_\Sone  = \beta_\Sone^\top \beta_\Sone \frac{1-\rho}{N}$.

\subsection*{Proof of Theorem \ref{thm:lcs}}

We first show that $\SNR = \log D$ is a necessary condition. For any $D$, $K$ or $\SNR$ assume $N$ is much larger such that
\[ E \left[ \ln \left( 1 + (1-\rho)\frac{\beta_\Sone^\top \beta_\Sone \SNR}{N} \right) \right] \asymp E \left[(1-\rho)\frac{\beta_\Sone^\top \beta_\Sone \SNR}{N} \right] = (1-\rho)\frac{i \sigma^2 \SNR}{N}. \]
Then the necessary condition given by Theorem 3.1 is
\[ N > C \max_i \frac{\log\binom{D-K}{i}\binom{K}{i}}{(1-\rho)\frac{i \sigma^2 \SNR}{N}} \]
which readily leads to the condition that
\begin{equation} \label{eq:SNR}
\SNR > C \max_i \frac{\log\binom{D-K}{i}\binom{K}{i}}{(1-\rho) i \sigma^2} \asymp \log D
\end{equation} 
for $\sigma$ constant.

From the upper bound given by Theorem 3.1, the sufficiency bound in Theorem 3.2 is obtained in a straightforward manner, by looking at conditions where $N f(\rho)$ goes to infinity. So for each $i$, we have
\[ P(E_i) \leq 2^{-\left( N \frac{1}{2} \log \left( 1 + (1-\rho) \frac{2 i \sigma^2 \SNR}{N}\right) - \frac{i}{4} \log 4 - \log\binom{D-K}{i}\binom{K}{i} \right)}, \]
then, as $\log\binom{D-K}{i}\binom{K}{i} = \Theta(i \log (D/i))$ dominates $\frac{i}{4} \log 4$ we can see that the following is a sufficient condition on $N$ for exact support recovery:
\begin{equation} \label{eq:cs_suff}
N > (1+\epsilon) \max_{i=1,\ldots,K} \frac{2\log\binom{D-K}{i}\binom{K}{i}}{\log \left( 1 + (1-\rho) \frac{2i \SNR \sigma^2}{N} \right)}. 
\end{equation}

Assume $\SNR > C\, \frac{\log D}{(1-\rho)\sigma^2}$. 
Also assume $N = \Omega\left(\frac{K \log (D/K)}{\log(1+(1-\rho)\sigma^2)}\right)$, as in the theorem statement. Then, the bound in \eqref{eq:cs_suff} becomes
\[ \max_{i=1,\ldots,K} \frac{2\log\binom{D-K}{i}\binom{K}{i}}{\log \left( 1 + (1-\rho) \frac{2i \SNR \sigma^2}{N} \right)} \asymp \max_{i=1,\ldots,K} \frac{i \log(D/i)}{\log\left( 1 + 2C\frac{i}{K}\log(1 + (1-\rho)\sigma^2)\frac{\log D}{\log (D/K)}\right)}, \]
where we assume $\sigma^2$ constant, w.l.o.g., since the scaling of elements of $\beta_S$ can instead be incorporated into $\SNR$ to obtain an equivalent model as we did in the proof of Theorem 3.1.

First, consider the case $K = o(D)$. Then the sufficient condition reduces to
\[ N > \max_{i=1,\ldots,K} \frac{i \log D}{\log\left( 1 + 2C\frac{i}{K}\log(1 + (1-\rho)\sigma^2)\right)}, \]
which, for the case $i = o(K)$ is
\[ N > \frac{i \log D}{C \frac{i}{K}\left(\log(1 + (1-\rho)\sigma^2)\right)} \asymp \frac{K \log (D/K)}{\log(1 + (1-\rho)\sigma^2)}, \]
which is satisfied for chosen $N$. For $i = \Theta(K)$, asymptotically, we have
\[ N > \frac{K \log D}{\log \left(1 + 2C\log(1 + (1-\rho)\sigma^2)\right)} \asymp \frac{K \log (D/K)}{\log \left(1 + \log(1 + (1-\rho)\sigma^2)\right)},\]
which is also satisfied by $N$.

Second, consider the case $K = \Theta(D)$. We then have the condition
\[ N > \max_{i=1,\ldots,K} \frac{i \log (D/i)}{\log\left( 1 + 2C\frac{i}{K}\log(1 + (1-\rho)\sigma^2) \log D\right)}, \]
which for $i = o(K)$, is asymptotically equivalent to
\[ N > \frac{i\log D}{2C \frac{i}{K} \log(1 + (1-\rho)\sigma^2) \log D} = \frac{K}{2C \log(1 + (1-\rho)\sigma^2)} \asymp \frac{K \log(D/K)}{\log(1 + (1-\rho)\sigma^2)} \]
which is satisfied for chosen $N$. For $i = \Theta(K)$, asymptotically we have the condition
\[ N > \frac{K \log (D/K)}{\log\left( 1 + \log(1 + (1-\rho)\sigma^2)\log D \right)}, \]
which is also satisfied for chosen $N$.

The necessity bound is obtained by using the derived mutual information expression and looking at the case $i=K$. From Lemma 3.1, we have
\[ I(X_\Sone ;Y |  X_\Stwo, \beta_S, \mu) \asymp E_{\beta_\Sone} \left[ \log \left( 1 + (1-\rho) \frac{\beta_\Sone^\top \beta_\Sone \SNR}{N} \right) \right], \] 
which leads to the following necessary condition, as given by Theorem 2.2:
\[ N \geq \max_{i=1,\ldots,K} \frac{\log\binom{D-K+i}{i}}{E_{\beta_\Sone} \left[ \log \left( 1 + (1-\rho) \frac{\beta_\Sone^\top \beta_\Sone \SNR}{N} \right) \right]}. \]

Note that $E_{\beta_\Sone} \left[ \log \left( 1 + (1-\rho) \frac{\beta_\Sone^\top \beta_\Sone \SNR}{N} \right) \right] \leq \log \left( E_{\beta_\Sone} \left[ 1 + (1-\rho) \frac{\beta_\Sone^\top \beta_\Sone \SNR}{N} \right] \right) = \log \left( 1 + (1-\rho) \frac{i \sigma^2 \SNR}{N} \right)$ due to Jensen's inequality, therefore the following is also a necessary condition, where we consider only $i=K$:
\begin{equation}\label{eq:cs_necessary}
N \geq  \frac{\log\binom{D}{K}}{\log \left( 1 + (1-\rho) \frac{K \sigma^2 \SNR}{N} \right)} \asymp \frac{K \log(D/K)}{\log \left( 1 + (1-\rho) \frac{K \sigma^2 \SNR}{N} \right)}.
\end{equation}

Assume $\SNR = \Theta(\log(D/K))$, which is given by \eqref{eq:SNR} for $\sigma^2 = O(1)$ and $i=K$. It is then clear that \eqref{eq:cs_necessary} does not hold for $N = o(K \log(D/K))$, since $\frac{K \log(D/K)}{N} \geq \log \left(1 + (1-\rho)\sigma^2 \frac{K \log(D/K)}{N} \right)$ for $\sigma^2 = O(1)$. However for $N = \Omega(K \log(D/K))$, the condition \eqref{eq:cs_necessary} is
\[ N = \Omega\left( \frac{K \log(D/K)}{\log \left( 1 + (1-\rho)\sigma^2 \right)} \right), \]
which proves the lower bound in Theorem 3.2.

\subsection*{Proof of Theorem \ref{thm:cs_additive_bound}}

To show the upper bound on error probability given in Theorem 3.3, we will write $P(Y | Z_\Sone, Z_\Stwo)$ and compute $E_o(\delta)$. For clarity, we consider fixed $\bS = \sigma$ as we did initially did in the proof of Theorem 3.1. Note that we can write
\[ P(Y | Z_\Sone, Z_\Stwo) = P(Y | Z_\Sone, Z_\Stwo, \mu) = \int_{\mathbb{R}^K} P(Y | X_\Sone, X_\Stwo) P(X_S | Z_S ,\mu) \dx X_S. \]

The first term is given by ${\cal N}(y - x^\top \bS; 0, 1/\SNR)$ as before, for $Y = y$ and $X_S = x$. Let $\alpha = \frac{1}{1+\nu}$, then using the conditional probability of jointly Gaussian random vectors, we have
\[ P(X_S = x | Z_S = z ,\mu) = {\cal N}\left(x; (1-\alpha) \mu 1_K + \alpha z, \frac{1-\rho}{N} (1-\alpha) I_K \right), \]
then, considering only sums of $X_S$ and $Z_S$ as $x$ and $z$ since $\bS = \sigma$, as we did in the proof of Theorem 3.1, the integral is
\begin{align*}
  P\left(Y=y \Big| \sum_{k \in S} Z_k=z, \mu\right) & = \int_{\mathbb{R}} {\cal N}(y - \sigma x; 0, 1/\SNR) \, {\cal N}\left(x; (1-\alpha) K \mu + \alpha z, \frac{1-\rho}{N} (1-\alpha) K \right) \dx x \\
  & = \frac{1}{2\pi}\sqrt{\frac{1}{AB}} \int_{\mathbb{R}} \exp \left( - \frac{(x-C)^2}{2A}\right) \exp \left( - \frac{(y - \sigma x)^2}{2B}\right) \dx x \\
  & = \frac{1}{2\pi}\sqrt{\frac{1}{AB}} \int_{\mathbb{R}} \exp \left( - \frac{(x-G)^2}{2\frac{AB}{A \sigma^2 + B}}\right) \exp \left( - \frac{(y - \sigma C)^2}{2(A \sigma^2 + B)}\right) \dx x \\
  & = \frac{1}{2\pi}\sqrt{\frac{1}{AB}} \exp \left( - \frac{(y - \sigma C)^2}{2(A \sigma^2 + B)}\right) \sqrt{\frac{AB}{2\pi(A \sigma^2 + B)}} \\
  & = \sqrt{\frac{1}{2\pi (A \sigma^2 + B)}} \exp \left( - \frac{(y - \sigma C)^2}{2(A \sigma^2 + B)}\right) \\
  & = {\cal N} \left(y - \alpha \sigma z - \alpha(1-\alpha)\sigma K \mu; 0, \frac{1}{\SNR} + \frac{(1-\rho)(1-\alpha)K\sigma^2}{N}\right),
\end{align*}
where $A = \frac{1-\rho}{N} (1-\alpha) K$, $B = \frac{1}{\SNR}$, $C = (1-\alpha) \mu K + \alpha z$ and $G = \frac{A \sigma y + B C}{\sqrt{A\sigma^2 + B}}$. The last equation follows through the steps used to show \eqref{eq:cs_pyx}. For the first equality, we compute and replace the probability distributions w.r.t.\ sums $x$ and $z$. We obtain the third equality by rewriting the terms inside the exponentials to obtain a square term with $x$. Then, we take the second exponential outside the integral and compute the integral, which gives us the fourth equality. Note that $G$ does not affect the integration result. Finally in the last step we note that the resulting expression is a Gaussian distribution with the given form.

Note that the resulting probability distribution is the same as $P(Y | X_S)$ we used in the proof of Theorem 3.1 except a few differences: $\sigma$ is replaced with $\alpha \sigma$, $\frac{1}{\SNR}$ is replaced with $\frac{1}{\SNR} + \frac{(1-\rho)(1-\alpha)K\sigma^2}{N}$ and lastly there is an extra $(1-\alpha) \mu K$ term. This last term does not affect the resulting lower bound on the error exponent $E_o$, since it disappears in the integration over $Y$ like the other $\mu$ terms. We also note that $P(Z_\Sone | \mu)$ and $P(Z_\Stwo | \mu)$ terms in the integral \eqref{eq:Eo_lb} are different than $P(X_\Sone | \mu)$ and $P(X_\Stwo | \mu)$. To account for this difference, we need to replace the variance $\frac{1-\rho}{N}$ with $\frac{(1-\rho)(1+\nu)}{N}$ in the integrations w.r.t.\ $z_1$ and $z_2$ that follow.

Finally, by doing the necessary replacements outlined above and following the proof of Theorem 3.1, we obtain the following error exponent for fixed $\bS = \sigma$:
\[ E_o(\delta) = \frac{\delta}{2} \log \left( 1 + \frac{1-\rho}{1+\nu} \frac{i \sigma^2 \SNR}{N (1+\delta)\xi} \right), \]
where $\xi = 1 + \frac{(1-\rho)\nu}{1+\nu} \frac{K \SNR \sigma^2}{N}$. Then the same analysis in the proof of Theorem 3.1 can be employed for random $\bS$, to obtain the lower bound, 
\[ E_o(1) \geq \frac{1}{2} \log \left( 1 + \frac{1-\rho}{1+\nu} \frac{2 i \sigma^2 \SNR}{N \xi} \right) - \frac{i}{4N} \log 4, \]
which proves the upper bound on error probability given in Theorem 3.3.

\subsection*{Proof of Theorem \ref{thm:cs_additive}}

We analyze the upper bound given in Theorem 3.3 to obtain the sufficient condition on $N$. First, note that $\frac{(1-\rho)\nu}{1+\nu} \leq 1$, therefore $\xi \leq 1 + \frac{K \SNR \sigma^2}{N}$.

Let $\SNR = c \log(D/K)$ for now, which is more relaxed than the $\SNR$ condition we assume in the theorem and assume $N = \Omega\left( \frac{K \log(D/K)}{\log\left( 1 + \frac{1-\rho}{1+\nu} \sigma^2 \right)} \right)$. Then it is easy to see that $\xi = O(1)$ for $\sigma^2 = O(1)$. As before, we can assume $\sigma^2 = O(1)$ w.l.o.g.\ since otherwise we can incorporate its scaling into $\SNR$. Then for some constant $C > 0$, we have the lower bound
\[ E_o(1) \geq \frac{1}{2} \log \left( 1 + \frac{1-\rho}{1+\nu} \frac{2 c i \sigma^2 \SNR}{C N} \right) - \frac{i}{4N} \log 4, \]
and therefore we have 
\[ P(E_i) \leq 2^{-\left( N \frac{1}{2} \log \left( 1 + c^\prime \frac{1-\rho}{1+\nu} \frac{i \sigma^2 \SNR}{N} \right) - \frac{i}{4} \log 4 - \log \binom{D-K}{i}\binom{K}{i} \right)}, \]
for a constant $c^\prime > 0$.

Following the arguments in the proof of Theorem 3.2, we can see that for $N$ chosen as above, $P(E)$ goes to zero, proving the theorem.

\bibliographystyle{unsrt}
\bibliography{references_dep}

\end{document}